\documentclass{article}

\usepackage{multicol}
\usepackage{caption}
\usepackage{longtable}
\usepackage[english]{babel}

\usepackage{mathrsfs}
\usepackage{multirow}
\usepackage{tikz}
\usetikzlibrary{quantikz2}
\date{}

\usepackage[letterpaper,top=2cm,bottom=2cm,left=1.5cm,right=1.5cm,marginparwidth=1.75cm]{geometry}
\usepackage{amsmath}
\usepackage{graphicx}
\usepackage[colorlinks=true, allcolors=blue]{hyperref}

\graphicspath{ {./Images/} } 
\title{\Large\bf{Towards Quantum Dynamics Simulation of Physical Systems: A Survey}}

\author{Rikteem Bhowmick{$^{1^{\dagger}}$},~ Navaneeth Krishnan Mohan{$^{1}$},~ Devesh Kumar{$^{1}$},\\~ Rohit Chaurasiya{$^{2}$},~ Nixon Patel{$^{1}$}\\\vspace{-8pt}{\small~}\\
	{$^{1}$}Qulabs Software (India) Pvt. Ltd;\\ {$^{2}$}Dept. of Applied Physics, Delhi Technological University, New Delhi, India;\\ 
	{\small{{$^{\dagger}$}Corresponding to: \tt{bhowmickrikteem@gmail.com}}}
}

\date{March 23, 2023}
\begin{document}
	\maketitle
	\begin{abstract}
		After the emergence of quantum mechanics and realising its need for an accurate understanding of physical systems, numerical methods were being used to undergo quantum mechanical treatment. With increasing system correlations and size, numerical methods fell rather inefficient and there was a need to simulate quantum mechanical phenomena on actual quantum computing hardware. Now, with noisy quantum computing machines that have been built and made available to use, realising quantum simulations are edging towards a practical reality. In this paper, we talk about the progress that has been made in the field of quantum simulations by actual quantum computing hardware and talk about some very fascinating fields where it has expanded its branches too. Not only that, but we also review different software tool-sets available to date which are to lay the foundation for realising quantum simulations in a much more comprehensive manner.
	\end{abstract}
\begin{multicols}{2}
	\section{Introduction}
	The beginning of the study of quantum mechanics dates back to 19th century with Young's double slit experiment and the seminal work on Black body radiation by Gustav Kirchhoff and Ludwig Boltzmann in 1862. Max Planck around 1900 suggested the quantization of electromagnetic energy to explain Black Body radiation and Albert Einstein in 1905 postulated that light is made of individual photons quoting Planck's hypothesis to explain the Photoelectric effect. In 1926, Erwin Schrödinger formulates the wave equation which laid the foundation of our understanding of the dynamics of a quantum system. \\
	To explain the dynamics of an atom coupled to an electromagnetic field, the \textit{master equation} was formulated governing the equation of motion for reduced density operator of an atom interacting with multi-mode EM field thus building the foundation of the study of quantum mechanical light-matter interaction.
	Ever since consideration of quantum mechanical treatment became significant, numerical methods of solving differential equations were devised when analytical methods fell inefficient and were difficult to solve. Moreover, with increasing system size and variables required to solve a quantum system accurately, numerical methods were either inefficient or classical computing machines (HPCs and Supercomputers) were incapable to handle the large computations required which is especially true when there is a strong correlation between system parties\cite{tacchino2020quantum}. 
	In 1982, Richard P. Feynman described his realisation that in order to replicate the dynamics of a quantum system, the inherent computing machine must work on the quantum mechanical laws, only then one can accurately simulate the dynamical evolution of any quantum system with appropriate resource utilisation. 
	
	In this paper, we have explored how in recent years quantum simulations are progressing towards reality and how recent realisations have been achieved to mimic the behavior of quantum systems with available universal programmable quantum computing devices as well as specific purpose quantum computing machines. We have also discussed a few areas where in recent years quantum simulations have drastically changed our understanding and perspective. We have structured our paper in the following manner- 
	in Section \ref{2} we talk about Quantum simulation for simulating quantum systems and processes and describe the different classes of quantum simulation namely, \textit{digital}, \textit{analog}, and \textit{digital-analog simulations}. Section \ref{3} discusses various transformation techniques used for a physical system to qubit mapping. In Section \ref{4}, we discuss different algorithms for Hamiltonian simulations. Section \ref{5} talks about various areas of study for quantum simulation and showing quantum advantage. In Section \ref{6}, we discuss about various Error mitigation techniques for mitigating errors arising from digital quantum simulations. The final Section \ref{7} discusses various software tool sets available to perform quantum simulations and narrates the current state-of-the-art towards achieving quantum dynamics simulation.
	
\section{Quantum simulation}\label{2}
Quantum simulations can be used to simulate the behaviour of other quantum systems efficiently and mimic processes (tunneling, quenching,
molecular resonance, lasing, fluorescence, etc.) which are difficult to solve because of the Quantum mechanical interactions (like superposition and entanglement) among the system constituents. This is why the domain of Quantum Simulation is one of the emerging fields of research since the last century. In recent years, the interest towards quantum simulations has grown rapidly as this area of research finds applications in multiple industry-driven sectors like the semiconductor and material designing industry, simulating condensed matter physics problems, light-matter interaction models to open quantum systems, simulating problems in quantum chemistry, drug discovery, etc.

The evolution of the Quantum system is governed by Schrodinger's equation, hence the core problem of the quantum simulation is solving Schrodinger's equation under different physical conditions.
The analytical approach can provide complete insights about the system of interest under some approximation but is not capable of handling large systems as the physical (or geometrical) interpretation of the low-level interaction becomes less intuitive with the increase in the number of particles. Scientists have suggested many theoretical methods (i.e Variational method, Perturbation theory, WKB approximation, 
Hartree-Fock method, Density functional theory, etc.), but none of these methods can provide the solution accurately regardless of the choice of quantum systems. Once the analytical method of solving equations became difficult to solve, scientists started focusing on numerical 
methods for Quantum simulation. Since last 5-6 decades researchers are extensively using efficient numerical algorithms (Quantum Monte Carlo, 
Crank-Nicolson algorithm, direct discretization of time evolution operator, matrix method, etc.) and high-performance computing clusters for performing large-scale quantum simulations. Since last decade, 
due to the emergence of artificial intelligence, a few groups of researchers have started exploring the machine learning methods for quantum simulations 
(NetKet and other examples).

Though numerical simulation and machine learning stand out to be efficient alternatives (under rigorous approximation) for quantum simulation, 
large scale practical quantum simulations are still hard to achieve because of the exponential requirement of time and computational memory. Imitating the plethora of quantum mechanical properties and processes of nature demands fundamental quantum mechanical treatment, and simulation of such a system can be accurately performed by a computing machine that inherently obeys quantum mechanical laws and can easily take into account the strong correlations of a quantum mechanical system\cite{feynman2018simulating}. Moreover, quantum systems can store large amounts of information in a relatively small amount of physical space\cite{georgescu2014quantum}. One can perform quantum simulation digitally or analogically by utilizing an inherent quantum system. In the following subsection, we discuss these methods in detail.

\subsection{Digital Quantum Simulation}\label{DQS}
Digital quantum simulation is the process of simulating the dynamics of a physical system by equivalent quantum circuits under controlled conditions. The idea fundamentally encapsulates the notion of a programmable quantum computer which is capable of simulating a multitude of physical systems and processes hence also called as Universal Quantum Simulation. To simulate the dynamics of a target quantum system, the hamiltonian of the target system containing all the information characterizing the system \cite{tacchino} is transformed and encoded to a quantum register native to a quantum hardware and translated into sequence of gate operations \cite{tacchino}. Initialising the quantum circuit with an appropriate initial state, the circuit is evolved in time in digital steps through a sequence of unitary operations recreating the dynamics of the physical system \cite{nielsen2002quantum} as well as its physical properties.
Any digital quantum simulation can be performed in five steps \cite{tacchino}- \\
Step 1: Defining the model hamiltonian $\mathcal{H}$ of a physical system containing all the fundamental interactions. \\
Step 2: Mapping the problem hamiltonian $\mathcal{H}$ onto qubit pauli algebra $\mathcal{H}\rightarrow H({\sigma_a})$ by mathematical transformations.\\ 
Step 3: Decomposing the hamiltonian onto a sum of local interaction hamiltonians $H\rightarrow\sum_i H_i$ and the simulation of such hamiltonians will be discussed in section \ref{4}.\\ 
Step 4: Now, each local unitary $\exp(-iH_it)$ or, $\exp(-iH_it/n)$ are translated to sequence of quantum gates by mathematical identities.\\
Step 5: Finally the circuit is initialised with an appropriate initial state to extract a desired observable by appending appropriate measurement at the end of the circuit.\\
A digital quantum simulator is quite similar to a quantum computer but still different in how its made use of. A quantum computer can be used to implement a variety of quantum algorithms while a digital quantum simulator can be used for simulating optimisation problems, thermodynamics and time evolution of physical systems and processes \cite{sanchez2018quantum}. Now, being similar to a quantum computer, a digital quantum simulator also surrounds from the problem of qubit decoherence, hence finds a challenge with implementation on a real hardware. Application of DQS doesn't solely revolves around simulating temporal evolution of system but can be used for obtaining partition functions, phase estimation for computing eigenvalues of Hamiltonian, finding ground state, etc. 

\subsection{Analog Quantum Simulation}
A quantum computer is not a necessity for performing quantum simulations, simpler problem specific quantum devices capable of emulating a quantum mechanical process can also be used to perform quantum simulations. These is called as Analog quantum simulations \cite{georgescu2014quantum}. Unlike digital quantum simulations, analog quantum simulations are problem specific and find a narrow range of physical quantum processes they can simulate. Thus, analog quantum simulation schemes are not universal but they are comparatively more accurate as they are designed to emulate some specific quantum systems. The system hamiltonian is mapped to the simulator hamiltonian via an operator A such that the mapping is reversible \cite{georgescu2014quantum},
$$\ket{\psi(0)}=A\ket{\phi(0)}\hspace{3mm}\text{such that},\hspace{3mm}\ket{\phi(t)}=A^{-1}\ket{\psi(t)}$$
where, $\ket{\psi(0)}$ is the simulator state and $\ket{\phi(0)}$ is the system state. The mapping is challenging to find at times and may involve additional external fields or ancillary systems to mediate various interactions \cite{georgescu2014quantum}.Also, quantum error correction can be easily implemented with DQS schemes, however analog simulations being special-purpose hardware models, error correction is challenging to incorporate \cite{sarovar2017reliability}.  

\subsection{Digital-Analog Quantum Simulation}
Analog quantum simulations are scalable as they allow simulations with larger number of particles \cite{lamata2018digital}, while digital quantum simulations allow to simulate various kinds of possible interactions the system is subject to. Combining this easier scalability of analog approaches with universality and flexibility of digital quantum simulation method, gives rise to Digital-Analog quantum simulation \cite{tacchino2020quantum}. The number of variables, digital quantum circuits needs to simulate are reduced fairly with integration of analog system as it allows direct simulation of a quantum system. Moreover, the required number of gates by the digital quantum simulation is also reduced hence reducing the experimental error. Digital-Analog quantum simulations of Rabi \cite{lamata2018digital} and Dicke model \cite{lamata2018digital} along with fermion-fermion scattering \cite{lamata14} has been performed.

\section{Transformation and Mapping schemes}\label{3}
This section covers some of the most common transformations in quantum simulation. Because quantum computers rely on qubit-based systems that obey Pauli algebra, transformations are crucial. The operators of a system with a different operator algebra must be translated to the Pauli-algebra if it is to be simulated on a quantum computer. In this section, we'll look at three different transformations. Holstein–Primakoff transformation, Jordan-Wigner transformation, Bravyi-Kitaev transformation.
\subsection{Jordan-Wigner transformation}

It is really important to have an understanding of Jardan-Wigner Transformation inorder to create algorithm using Digital Quantum Simulation. This transformation can be used to map the system of spin-$\frac{1}{2}$ particles, that are qubits to the system of Fermions. Fermions are those particles which have odd half-integer spin. Quarks and Leptons are examples of fermions. If we consider a 1-D array of spin-$\frac{1}{2}$ particles then at each site either the spin are in ($\ket{\uparrow}$) or down state ($\ket{\downarrow}$). But if we have a 1-D array of fermions then at each site there is either a fermion or no fermion. 

Let us consider of a system of n-qubits with usual state space $C^{2^n}$. Let $X_j$, $Y_j$ and $Z_j$ be the Pauli operators which will be acting on the $j^{th}$ qubit. With help of Jordan-Wigner we can define the fermionic operators, $\hat{a_j}$ and $\hat{a_{j}^{\dagger}}$ in terms of Pauli operators. The action of the operator $\hat{a_{j}}$ applied to the number representation of occupation state $\ket{\alpha}$ is: 
\begin{enumerate}
	\item If $\alpha_j = 0$, then $\hat{a_{j}}\ket{\alpha}=0$
	\item If $\alpha_j = 1$, then considering $\alpha_{\prime}$ to be a vector resulting after the $j_{th}$ entry is changed to 0, we have $\hat{a_{j}}|\alpha\rangle=-(-1)^{s_{\alpha}^{j}}\left|\alpha^{\prime}\right\rangle$. where s is $s_{\alpha}^{j} \equiv \sum_{k=1}^{j-1} \alpha_{k}$
	
\end{enumerate}

\begin{equation}\label{JW transformation}
	\hat{a_{j}} \equiv-\left(\otimes_{k=1}^{j-1} Z_{k}\right)\otimes\hat{\sigma_j}
\end{equation}

The Jordan-Wigner transform can be given as shown in eq.\ref{JW transformation} if the computational basis is considered to be the occupation number state. This transformation can be written otherway round, meaning that the Fermionic operators can be used to express the Pauli-operators. 

\begin{equation}\label{Z_j}
	\hat{X_{j}} = \hat{a_{j}}\hat{a_j}^{\dagger}-\hat{a_j}^{\dagger}\hat{a_{j}}
\end{equation}

For Pauli-Z operator one can write the expression as in eq \ref{Z_{j}}. Using the fact that the $\hat{X}=\hat{\sigma_{j}} +\hat{\sigma_{j}^{\dagger}}$, one can also express $\hat{X_{j}}$ in terms of Fermionic operators as shown in eq.\ref{X_{j}} .

\begin{equation}\label{X_j}
	\hat{X_{j}} = -(Z_1...Z_{j-1})(\hat{a_j}+\hat{a_j}^{\dagger})
\end{equation}
In same way we can express $Y_j$ as in equation \ref{Y_j}.

\begin{equation}\label{Y_j}
	Y_{j} = -\iota(Z_1...Z_{j-1})(\hat{a_j}^{\dagger}-\hat{a_j}) 
\end{equation}

The more important to the study of digital quantum simulation is the product form of the Pauli operators. These product form allows one to describe XX, YY, ZZ interarctions. Some of them have been given in the eq.\ref{product}

\begin{equation}\label{product}
	\begin{aligned}
		Z_{j}&=\hat{a_j} \hat{a_j}^{\dagger}-\hat{a_j}^{\dagger} \hat{a_j} \\
		X_{j} X_{j+1} &=\left(\hat{a_j}^{\dagger}-\hat{a_{j}}\right)\left(\hat{a}_{j+1}+\hat{a}_{j+1}^{\dagger}\right)\\
		Y_{j} Y_{j+1}&=-\left(\hat{a_{j}}^{\dagger}+\hat{a_{j}}\right)\left(\hat{a_{j+1}}^{\dagger}-\hat{a_{j+1}}\right) \\
		X_{j} Y_{j+1}&=i\left(\hat{a}_{j}^{\dagger}-\hat{a_{j}}\right)\left(\hat{a_{j+1}}^{\dagger}-\hat{a_{j+1}}\right) \\
		Y_{j} X_{j+1}&=i\left(\hat{a_{j}}^{\dagger}+\hat{a_{j}}\right)\left(\hat{a}_{j+1}^{\dagger}+\hat{a}_{j+1}\right)
	\end{aligned}
\end{equation}

\subsection{Holstein-Primakoff transformation}

Yet another important transformation is the Holstein-Primakoff transformation. This allows the treatment of spin operators in the form of bosonic ones. This transformation maps the spin-S moments to the bosonic creator and annihilation operators as in eq. \ref{HPtrans}. Hence allowing us to bosonize the spin systems. 

\begin{equation}\label{HPtrans}
	\begin{aligned}
		\hat{S}_{j}^{z} &=S-\hat{n}_{j} \\
		\hat{S}_{j}^{+} &=\sqrt{2 S-\hat{n}_{j}} \hat{b}_{j} \\
		\hat{S}_{j}^{-} &=\hat{b}_{j}^{\dagger} \sqrt{2 S-\hat{n}_{j}}
	\end{aligned}
\end{equation}

In the above equaiton, $\hat{b^{\dagger}}$ and $\hat{b}$ are the bosonic creation and annihilation operators. These satisfy the bosonic commutation relations as shown in eq.\ref{commutation} while n is the number operator defined as $\hat{n}=\hat{b}\hat{b^{\dagger}}$.

\begin{equation}\label{commutation}
	\begin{aligned}
		&{\left[\hat{b}_{i}, \hat{b}^{\dagger}_{j}\right]=\delta_{i j}} \\
		&{\left[\hat{b}_{i}, \hat{b}_{j}\right]=\left[\hat{b}_{i}^{\dagger}, \hat{b}_{j}^{\dagger}\right]=0}
	\end{aligned}
\end{equation}

At thermal energy lower than exchange energy ($k_BT << J$), the above transformation can be approximated as shown in eq.\ref{approx}

\begin{equation}\label{approx}
	\begin{aligned}
		&\hat{S}_{j}^{z}=S-\hat{n}_{j} \\
		&\hat{S}_{j}^{+} \approx \hat{b}_{j} \sqrt{2 S} \\
		&\hat{S}_{j}^{-} \approx \hat{b}_{j}^{\dagger} \sqrt{2 S}
	\end{aligned}
\end{equation}

This allows spinorisation of bosonic particles.

\begin{equation}
	\hat{b^{\dagger}} \hat{b}=S_1+S_{z} \quad \hat{b^{\dagger}}=\hat{S_{+}} \frac{1}{\sqrt{S_1-S_{z}}} \quad \hat{b}=\frac{1}{\sqrt{S_1-S_{z}}} \hat{S_{-}}
\end{equation}

With this transformation one can easily embed the bosonic system to spin qubits for particular hardware, further helping in digital simulation on quantum computers.

\subsection{Bravyi-Kitaev Transformation}\label{BK Transformation}

The Bravyi–Kitaev transformation~\cite{bravyi2002fermionic} is an alternative to the Jordan–Wigner transformation~\cite{jordan1993paulische}, which can be used for mapping fermions to spins. This transformation was defined in~\cite{bravyi2002fermionic} under the context of quantum computing using fermions. Authors in~\cite{tranter2015b,seeley2012bravyi} provide a detailed discussion of this mapping.

To mimic fermionic operators using qubits, two forms of information are required~\cite{seeley2012bravyi}: the occupation of the target orbital and the parity of the set of orbitals with index smaller than the target orbital. A matrix $\beta_n$ that works on bit strings corresponding to occupancy number basis vectors of length n can be used to represent the Bravyi-Kitaev basis encoding. The map from occupation number to this basis is as follows:
\begin{equation}
	b_i = \sum_j [\beta_n]_{ij} f_i   
\end{equation}
For an arbitrary index $j$, the Bravyi-Kitaev transformation involves three sets of qubits: the \emph{parity set} (qubits in the Bravyi-Kitaev basis that store the parity of all orbitals with index less than $j$), the \emph{update set} (qubits that store a partial sum including orbital $j$), and the \emph{flip set} (the qubits that determine whether qubit $j$ has the same parity or inverted parity with respect to orbital $j$). These index sets enables the explicit construction of the fermionic creation and annihilation operations in the Bravyi-Kitaev basis.\\
\\
\emph{Pauli operator representation of the fermionic creation and annihilation operators acting on orbital j in the Bravyi-Kitaev basis~\cite{seeley2012bravyi}:}
\begin{itemize}
	\item For even $j$: To represent $\hat{a}$ and $\hat{a^\dagger}$ in Bravyi-Kitaev basis for even $j$, $\hat{\sigma_z}$ is applied on all qubits in $P(j)$ (the parity set), $\hat{Q}^\pm$ (the qubit creation and annihilation operators in the occupation number basis) on qubit $j$, and $\hat{\sigma_x}$ on all qubits in $U(j)$ (the update set):
	\begin{equation}
		\begin{aligned}
			\begin{split}
				\hat{a}_j^\dagger &= X_{U(j)} \otimes \hat{Q}_j^+ \otimes Z_{P(j)} = \frac{1}{2} \big( X_{U(j)} \otimes X_{j} \otimes\\
				&Z_{P(j)}- i X_{U(j)} \otimes Y_{j} \otimes Z_{P(j)} \big) \\
				\hat{a}_j &= X_{U(j)} \otimes \hat{Q}_j^- \otimes Z_{P(j)} = \frac{1}{2} \big( X_{U(j)} \otimes X_{j} \otimes\\
				&Z_{P(j)}+ i X_{U(j)} \otimes Y_{j} \otimes Z_{P(j)} \big)
			\end{split}
		\end{aligned}
	\end{equation}
	
	\item For odd $j$: To represent $\hat{a}$ and $\hat{a^\dagger}$ in Bravyi-Kitaev basis for odd $j$, $\hat{\sigma_z}$ is applied on all qubits in $R(j)$ (the remainder set), $\hat{\Pi}^\pm$ (the qubit creation and annihilation operators that depends on the parity of the set of qubits in the flip set $F(j)$ ) on qubit $j$, and $\hat{\sigma_x}$ on all qubits in $U(j)$:
	\begin{equation}
		\begin{aligned}
			\begin{split}
				\hat{a}_j^\dagger &= X_{U(j)} \otimes \hat{\Pi}_j^+ \otimes Z_{R(j)} = \frac{1}{2} \big( X_{U(j)} \otimes X_{j} \otimes\\
				&Z_{P(j)}- i X_{U(j)} \otimes Y_{j} \otimes Z_{R(j)} \big) \\ 
				\hat{a}_j &= X_{U(j)} \otimes \hat{\Pi}_j^- \otimes Z_{R(j)} = \frac{1}{2} \big( X_{U(j)} \otimes X_{j} \otimes\\
				&Z_{P(j)}+ i X_{U(j)} \otimes Y_{j} \otimes Z_{R(j)} \big)
			\end{split}
		\end{aligned}
	\end{equation}
	Where $R(j)$, the remainder set ($\equiv P(j) \backslash F(j)$) is the set of qubits that are in $P(j)$ but not in $F(j)$.
\end{itemize}
These are useful fundamental findings. Using these results, other operators that can be regarded as products of these creation and annihilation operators may be turned into Bravyi-Kitaev basis. Authors in~\cite{seeley2012bravyi} have described several instances of such operator transformations, such as the number operator and the coulomb/exchange operator etc.

\section{State Preparation}
In order to extract a physical parameter or any desired physical observable, the quantum circuit intended for mimicking a particular quantum mechanical process needs to be initialised with an appropriate initial state, $\ket{\psi(0)}$.
State preparation methods aim to build a circuit that creates a state with as much overlap as possible with the target eigenstate. State preparation is critical in quantum chemistry and material simulation. In condensed matter, current efforts are focused on preparing ground or low-lying energy and the thermal state of many-body systems. While in quantum chemistry many work has been done by preparing the initial wavefunctions. These wavefunctions are approximations to the real wave-functions. For this configuration methods are used to find such wavefunction and are then prepared as initial states. These states could be used to extract the system's static properties (e.g., ground state energy) or as an initial state in the dynamic simulations. In-state preparation methods often  start from an easily preparable product state and apply some unitary transformations such that the output state has maximum fidelity with the target state. In the following sections, we will briefly discuss the ground, excited and thermal state preparation in Condensed Matter Physics. A large chunk of current works in this direction is focused on quantum chemistry but will be helpful in the context of quantum material simulation.
\subsubsection*{Ground State Preparation}
There are different approaches for ground state preparation. The first approach is to use Quantum Phase Estimation (QPE). It is the direct route to prepare the state but is very hard to implement in the NISQ era. Another method is to evolve the system from a simple initial state to the desired state either in real-time (\emph{Adiabatic State Preparation (ASP)}) or in imaginary time (\emph{Quantum Imaginary Time Evolution (QITE)}). But these methods are not efficient for every Hamiltonian. The favourite state preparation approach in the NISQ era is using Variational Quantum Algorithms. Using Variational Quantum Eigensolver (VQE), one obtains the ground state and the energy of a system. A hybrid quantum-classical approach is used in these algorithms, making it suitable for NISQ computers having a short coherence time. 

\emph{Quantum Phase Estimation}\label{QPE}: QPE \cite{kitaev95} is an essential sub-routine used in many quantum algorithms. Given a Hamiltonian, QPE can be used to calculate the energy eigenvalues. In the QPE algorithm, we use two registers: the first register (\emph{State register}) will be initialized in a state having ample overlap with the desired eigenstate and the second register (\emph{Ancilla register})in an equal superposition of the computational basis state. The QPE exploits the phase kickback effect to write the phase of Hamiltonian on the qubits in the ancilla register \cite{qpe}. The energy eigenvalues in the binary format could be retrieved by measuring ancilla qubits after performing an inverse quantum Fourier transform on them. At the end of the algorithm, the state register collapses to an eigenstate corresponding to the measured eigenvalue. In our case, the state register is made to collapse to the ground state at the end of QPE. Thereby QPE provides a way to construct the ground state. Now the efficiency of this method depends directly proportional to overlap between the initial state (of the state register) and the ground state. However, preparing the initial state with ample overlap with the ground state is not trivial. For the final state of the state register to have high fidelity with the desired ground state, one must repeatedly perform QPE. As a result, computational cost scales inversely with the amount of initial and ground state overlap. It is inefficient to use this method in NISQ devices. Nevertheless, QPE has been demonstrated in spin chains \cite{cruz20,li11}.

\emph{Adiabatic State Preparation}: In ASP \cite{guzik2005,wan22} the qubits will be initialized in the ground state of a simple Hamiltonian (easily preparable) and then adiabatically vary the Hamiltonian into the Hamiltonian of the target ground state. The Adiabatic Theorem powers adiabatic state preparation. Assume that we have a time-varying Hamiltonian $\hat{H}(t)$ initially $\hat{H}_{I}$ at t = 0 and $\hat{H}_{F}$ at some later time $t = t_{F}$. Suppose the system is initially in the ground state of $\hat{H}_{I}$ and the Hamiltonian's time evolution is sufficiently slow. In that case, the state is likely to remain in the ground state throughout the evolution, thus being in the ground state of $\hat{H}_{F}$ at $t = t_{F}$. In ASP, we mimick the adiabatic evolution using quantum gates. In theory, the evolution must be infinitesimally slow evolution time $T\rightarrow \infty$. But in practice, the $T$ should be at least $\mathcal O(\dfrac{1}{g_{min}^{2}})$ and here $g_{min}$ denotes the spectral gap; the gap between energy levels. It ensures the system does not undergo the transition to excited states during the evolution. 

In condensed matter, there are some works that use ASP. In \cite{wecker15}, they present the scheme for preparing initial states from arbitrary slater determinants and adiabatically evolve them to the ground state of the Hubbard model. Similarly, the ground-state phase diagram of the XY model is computed using ASP in \cite{francis22}.

\emph{Quantum Imaginary Time Evolution}: QITE works based on the concept of imaginary time evolution (ITE), a method used to determine the nearly exact ground-states in numerical simulation. As the name implies, we replace real-time with imaginary time in the evolution of this method. In the real-time evolution, the state of the system at the time $t$ is given by:
\begin{equation}
	\Psi (t)=\sum_{n}\exp(-iE_{n}t/\hbar)\psi_{n}
	\label{naveq-5}
\end{equation}
where $\psi_{n}$ represents the eigenstate of the system. Now in ITE, we replaces $t$ with imaginary time $i\beta$. Thus the state will be:
\begin{equation}
	\Psi (t)=\sum_{n}\exp(-E_{n}\beta/\hbar)\psi_{n}
	\label{naveq-6}
\end{equation}
Now in \eqref{naveq-5}, $\Psi (t)$ is the superposition of eigenstates $\psi_{n}$ oscillating with a frequency $E_{n}/\hbar $ that interfere to give final output state. Whereas in \eqref{naveq-5}, it is a superposition of exponentially decaying energy eigenstates. Since the decay rate is $E_{n}/\hbar $ in the long imaginary time limit $\beta \rightarrow \infty$:
\begin{equation}
	\Psi (t)\sim \exp(-E_{0}\beta/\hbar)\psi_{0}
	\label{naveq-7}
\end{equation}
Since $\psi_{0}$ decays very slowly compared to other eigenstates. Thus performing imaginary time evolution over the long imaginary time limit system will go to the ground state. We need to realize the non-unitary time evolution operator, $\mathcal{U}=\exp (-\beta \hat{H})$ (where $\hbar=1$), to simulate the imaginary time evolution. Numerical methods to perform ITE is discussed in \cite{brouwer05}. he quantum algorithm for performing imaginary time evolution-Quantum Imaginary Time Evolution (QITE) Algorithm -was recently proposed \cite{motta20}. The QITE algorithm uses approximate unitary updates to simulate the non-unitary imaginary-time evolution. QITE is an iterative algorithm in which we measure certain operators in the $n^{th}$ step and use those values to solve a system of linear equations that give the parameters for the unitary update in the $(n+1)^{th}$ time step.  Now, QITE converges to the ground state only if the initial state has a finite overlap with the ground state. Like ASP, the ground state of a many-body system can be prepared using the QITE algorithm without variational optimization. 

In condensed matter, QITE has been applied to find the ground state of the TFIM model \cite{motta20}. The QITE has also been used as a sub-routine in the simulation of scattering in the Ising model \cite{yeter21}. Shortly after QITE was proposed, a resource-efficient version of QITE was proposed, and they have used it to find the binding energies of certain molecules \cite{gomes20}.

\emph{Variational State Preparation and Variational Quantum  Eigensolver (VQE)}: This class of methods is considered to be useful for efficient ground state preparation in near-term devices. Here, one had to choose some ansatz for the ground state of Hamiltonian of interest. An ansatz consists of an initial state and a unitary circuit parametrized by some classical tunable parameters. First, we would calculate the ansatz's energy, which would give the upper bound of ground state energy. Then, using a classical optimizer, we try to minimize this energy by updating the parameters of the ansatz. Now repeating this hybrid classical-quantum loop, one could find the ground state energy, and the corresponding state will be the ground state. There are different versions of VQE that could be
used to access the lowest energy eigenstates. A detailed explanation of such algorithms and various ansatz is provided in \cite{cerezo21}. In addition, there also exists variational ansatz-based quantum simulation of imaginary time evolution \cite{mcardle19} that could be used to prepare the ground state.

When selecting ansatz state classes, many requirements must be met: on the one hand, the class must include an accurate approximation of the system's actual ground state. On the other side, one wishes for a class of circuits that can be readily implemented on a quantum computer, i.e. for a given set of accessible gates, qubit connectivity, and so on. There are many types of ansatzes. Some ansatzes that are being used in the Condensed Matter include; Hamiltonian Variational Ansatz \cite{martin22,wecker15b}, Hardware Efficient Ansatz \cite{kandala17}, tensor networks \cite{slattery21,liu19} and so on.
\subsubsection*{Excited State Preparation}
Most of the methods discussed in the ground state preparation could be extended to the excited state preparation. For instance, variational quantum deflation algorithms use the ground state prepared through VQE could be used to prepare the excited state and to calculate excited energies \cite{higgott19,jones19,sa22}. Another variational approach to obtaining low energy eigenstates is the sub-space expansion method \cite{mcclean17,colless18}. This method uses a sub-space of states generated from the estimated ground state to calculate the excited states. Similarly, an algorithm called the Qlanczos algorithm uses the measurement outcomes from the QITE algorithm to obtain all eigenstates of the system, including the excited state \cite{motta20}. Further, in principle, ASP can prepare the excited state if it can be connected to the initial state with a non-vanishing energy gap. The above algorithms compute total excited state energies, which must be subtracted from the ground state energy to get the system's excitation energies. Recently an algorithm called Quantum Equation of Motion was put forward that could be used to calculate the excited energies directly \cite{ollitrault20}.
\subsubsection*{Themal State Preparation}
State of the system which is at thermal equilibrium with the bath is called Thermal state or Gibbs state of the system. The corresponding density matrix of the system is given by:
\begin{equation}
	\rho_{G}= \frac{\exp(-\beta \hat{H})}{Z}
	\label{naveq-8}
\end{equation}
where $Z=tr(\exp(-\beta \hat{H}))$ and $\beta=\frac{1}{K_{B}T}$. The state \eqref{naveq-8} represents the equilibrium state of the system that is coupled to a very large Thermal bath. The size of the system is not essential; for example, it could be a single electron. But the bath is usually taken to be a macroscopic body. The Gibbs state will be a valid state of the system only if the bath is macroscopically large and complex. The equilibrium thermal Gibbs state is the most common state of quantum matter. Consequently, it has applications in quantum chemistry and materials research, high-energy physics, and computer science. Despite advancements in time evolution simulation, creating thermal states of quantum many-body systems remains a challenging but essential operation.

Early algorithms for the simulation of Gibbs states \cite{terhal2000} were based on the idea of coupling the system of interest to a set of ancillary qubits (``digital bath"). The system and bath qubits together are evolved under a joint Hamiltonian, thus mimicking the physical process of thermalization. Thermalization-based methods have two main disadvantages: they require additional qubits to describe the bath states and require time evolution under Hamiltonian for a long time $t$ for thermalization. Very recently, the digital bath approach was revamped using Random Quantum Circuits that have proven to be efficient in near-term quantum computers \cite{shtanko21}.

Later efforts use accurate techniques based on quantum phase estimation to execute computations in the Hamiltonian's eigenbasis \cite{poulin2009,riera12}. A ubiquitous example of such an algorithm is the Quantum Metropolis Algorithm \cite{temme2011,yung12}. Rather than generating the Gibbs state explicitly, it samples from it to obtain the equilibrium and static properties of the system. Another approach uses a modified QPE to get the Green's Function of the system and thereby the thermodynamic properties \cite{dallaire16}. Another algorithm called minimal effective Gibbs ansatz (MEGA) also calculates the Greens function of Micro-canonical ensembles using methods like QPE \cite{cohn20}. But rendering QPE is impractical for current noisy hardware.

The notion of imaginary time evolution is also used for thermal state preparation in Quantum computers. In \cite{chowdhury16}, they have discussed a method for preparing Gibbs state using an ancilla based Hamiltonian simulation in imaginary time. Alternatively, there exists an algorithm called quantum minimally entangled typical thermal state (METTS) algorithm that samples from the Gibbs state by applying the QITE algorithm on pure states \cite{sun21,motta20}. However, It is unclear how precise this approach is beyond short-range coupled systems, and its performance takes exponential time. In addition, the imaginary time evolution approach of the Gibbs state was recently demonstrated using Random Quantum Circuits in \cite{shtanko21}.

Several variational hybrid quantum-classical algorithms are there for approximate Gibbs state preparation. For instance, the Quantum Approximate Thermilization method uses the QAOA for thermal state preparations \cite{verdon19}. Further, there are variational approaches that use Thermo Field Double (TFD) states \cite{wu19,cottrell19, martyn19,sagastizabal21}, imaginary time evolution \cite{yuan19} and so on. On the other hand, variational algorithms do not offer a demonstrable benefit over classical computers, and their output state cannot be validated close to the thermal state \cite{mcclean18}.

\subsubsection*{Configuration Method for State Preparation}\label{section_conf}

This section specifically discusses state preparation methods specific to quantum chemistry. Most of the quantum chemistry algorithm use configurational methods for preparing the initial state. These configurational method rely on a good trial wave function consideration. The quantum algorithm proposed in \cite{spect_12aspuru2005simulated} promises to implement the quantum based FCI algorithm and solve for the system in polynomial time. It uses Hartree Fock (HF) wave function as the initial trial state. 

The research is ongoing for having to select the most suitable initial trial wave function. The authors in the paper \cite{spect_13wang2008quantum} calculated the energy eigenvalues of the multi-reference configuration interaction (MRCI) wave function using the Multiconfigurational self-consisted field (MCSCF)  wavefunction. These are different types of wavefunctions that can be used as the initial state in calculations of molecular properties like ground state and low-lying excited states.
The digital quantum simulation procedure involves a few steps. First, it is important to find the vibrational hamiltonian of the system. Mostly, the hamiltonian is described with the Born-Oppenheimer approximation and then it is diagonalised. This step is followed by mapping the hamiltonian into the type of qubit being used which depends on the kind of quantum computer used for performing the simulations. Efficient mapping methods are important as they decide the number of qubits used. Two famous methods of mapping are direct mapping and compact mapping. A space $\ket{s}$ is mapped as follows with n qubits as given in the eq.\ref{space-direct mapping}

\begin{equation}\label{space-direct mapping}
	\ket{s}=\otimes_{j=0}^{s-1}\ket{0}_{j}\ket{1}_{s} \otimes_{j=s+1}^{n-1}\ket{1}_{j}
\end{equation}

In this case, the annihilation operators take the form as shown in eq.\ref{annihilation op direct mapping}

\begin{equation}\label{annihilation op direct mapping}
	\hat{a}=\sum_{s=0}^{n-2} \sqrt{s+1}\ket{0}\bra{1}_{s}\otimes \ket{1}\bra{0}_{s+1}.
\end{equation}

In the case of compact mapping, the space is encoded into qubits as given in equation \ref{compact mapping}

\begin{equation}\label{compact mapping}
	\ket{s}=\left|b_{K-1}\right\rangle\left|b_{K-2}\right\rangle \ldots\left|b_{0}\right\rangle
\end{equation}
In the above equation, $\ket{s}$ is given in the binary representation. Also, the form of the creations operator is given as seen in eq.\ref{annihilation - compact}.
\begin{equation}\label{annihilation - compact}
	\hat{a}^{\dagger}=\sum_{s=0}^{n-2} \sqrt{s+1}|s+1\rangle\langle s|
\end{equation}
Compact mapping gives an advantage of using less number of qubits as compared to the direct mapping method. For direct mapping, the qubits required are $Mn$ while it is $Mlog(n)$ in the case of compact mapping. 

Once the mapping is done, the unitary is created from the hamiltonian and finally, a quantum circuit is created using the native gates. Using the digital quantum simulation method one can simulate Vibrational energy levels, Franck-Condon factors and Vibrational dynamics \cite{spect_6mcardle2019digital}.

\section{Hamiltonian Simulation}\label{4}
Hamiltonian simulation is a vital step in quantum simulation. As we discussed in the section \ref{DQS}, a DQS of a physical system proceeds through three stages, initial state preparation, Hamiltonian simulation, and readout of observables. The Hamiltonians we encounter during the simulation of physical systems belong to the class of local Hamiltonians. A local Hamiltonian $\hat{H}$ can be described as: 
\begin{equation} 
	\hat{H}=\sum_{j=1}^{L} \alpha_{j} h_{j}
	\label{naveqn-1}
\end{equation}
where $L=$ the number of terms in Hamiltonian, $h_{j}$ is the tensor product of pauli operators acting on $k$ out of $n$ qubits (hence ``$k$-local'') and $\alpha_{j}$ is the real coefficients. The aim of the Hamiltonian simulation is to implement the unitary time evolution operator $U=\exp(-i \hat{H} t)$ (for $\hbar=1$) in the quantum computer for a given time $t$ within an error bound $\epsilon$.

The local Hamiltonians belong to the class of much general Sparse Hamiltonian. In mathematics, a matrix is said to be sparse if most of its elements are zero. Thus a $d$-sparse Hamiltonian has at most $d$ non-zero elements in any row or column. These types of Hamiltonian are popular in quantum algorithms using quantum walks \cite{zhou21}. But in this article, we restrict our discussions to local Hamiltonian simulation in a programmable quantum computer. In the following sections, we will discuss the three most prevalent algorithms for Hamiltonian simulation.

\subsection{Product Formula Algorithm}

Product Formula (PF) is the earliest and most prevalent method for Hamiltonian simulation. This approach approximates the exponential of the sum of operators as the product of the exponential of individual operators. In the case of local Hamiltonians, if $[h_{i},h_{j}]=0$ then,
\begin{equation*}
	\resizebox{1\hsize}{!}{%
		$U=\exp(-i \hat{H} t)= \exp(-i \sum_{j}\alpha_{i}h_{j} t) = \prod_{j}  \exp(-it \alpha_{i}h_{j})$%
	}
\end{equation*}
But in general $[h_{i},h_{j}] \neq 0$ then the unitary evolution operator can be written as:
\begin{equation}
	\resizebox{1\hsize}{!}{%
		$U=\exp(-i \hat{H} t)= \left[\prod_{j}  \exp(-\frac{it}{r} \alpha_{i}h_{j} ) \right]^{r} + \mathcal{O}\left(\frac{(L\Lambda t)^{2}}{r}\right)$%
	}
\end{equation}
where $\Lambda= max_{j}\alpha_{j}$. This is called first order Lie-Suzuki-Trotter formula (Trotter formula). One can use higher orders of the Trotter formula to better approximate the unitary evolution operator. Thus the $(2k)^{th}$ order trotter formula $S_{2k}$ is given by: 
\begin{equation}
	\begin{split}
		S_{2k}(\Lambda) = &\left[S_{2k - 2}(p_{k}\Lambda)\right]^{2}S_{2k -2}((1- 4p_{k})\Lambda)\\
		&\left[S_{2k -2}(p_{k}\Lambda)\right]^{2}+\mathcal{O}\left(\frac{(L\Lambda t)^{2k+1}}{r^{2k}}\right)
	\end{split}
\end{equation}
where \resizebox{0.88\hsize}{!}{%
	$S_{2}(\Lambda)=\prod_{j=1}^{L}\exp(\alpha_{j}h_{j}\Lambda/2)\prod_{j=L}^{1}\exp(\alpha_{j}h_{j}\Lambda/2)$%
} 
\\
with $P_{k}=1/(4-4^{1/(2k-1)})\hspace{1mm}\text{for}\hspace{1mm}k>1$.

Approximation of evolution operator becomes better and better as we go to the higher-order trotter formula. But the asymptotic complexity of implementing higher-order trotter formula is very high. For instance, the run time of a $k^{th}$ order trotter formula is $\mathcal{O}(5^{k}tL\left(\frac{tL}{\epsilon}\right)^{1/k})$. Thus in a practical case, we don't go beyond $2^{nd}$ or $4^{th}$ order. In addition, gate count also increases with the number of terms in the Hamiltonian. This scaling becomes a problem while simulating Hamiltonians like electronic structure Hamiltonians, where the number of terms is large. Recently it was shown that one could use a random compiler than a deterministic compiler in trotterization for faster simulation \cite{campbell19}. This technique is named quantum stochastic drift protocol (qDRIFT). Random compiling also helps to reduce errors more than deterministic compiling.

\subsection{Taylor Series Algorithm}
The Taylor series algorithm uses a truncated Taylor series expansion of the time evolution operator to carry out the Hamiltonian simulation. This algorithm only applies to Hamiltonian that can be written as a linear combination of unitary operators. For instance, the local Hamiltonian we discussed. The time evolution is broken up into small intervals or segments. Within each time interval, the Taylor series expansion of the time evolution operator can be truncated into a small number of terms without increasing the error. The truncated Taylor expansion corresponds to a linear combination of unitary operators, which can be probabilistically implemented in a quantum computer using ancillary superpositions and controlled operations \cite{childs12}. In addition, we need to carry out a process called Oblivious Amplitude Amplification (OAA) for deterministic implementation of the time evolution (will discuss later). Thus at each time segment, time evolution can be implemented using the LCU operator method followed by OAA \cite{berry15}. This time evolution segment is repeated to carry out time evolution for the entire time duration $t$.  

\emph{Method}: We divide the evolution time into r intervals of $t/r$ duration. Within each interval, the Taylor expansion of time evolution operator is truncated to an order $K$.
\begin{equation}
	U=\exp(\frac{-it}{r}\hat{H}) \approx \sum_{k=0}^{K}\frac{1}{k!}(-i\hat{H}t/r)^{k}
	\label{naveqn-3}
\end{equation}
The order $K$ should be chosen appropriately so that each interval so that each interval utmost has an error $\epsilon/r$.  By substituting \eqref{naveqn-1} into \eqref{naveqn-3} and expanding the terms, we can see that $U$  can be expressed as a linear combination of unitary operators. It takes a form which we can generally represent as:
\begin{equation}
	\tilde{U}=\sum_{j}\beta_{j}V_{j}
	\label{naveqn-4}
\end{equation}
Where $\beta_{j}$ will be a constant greater than zero, and $V_{j}$ is a unitary operator. Here $V_{j}$ corresponds to the product of operators in the Taylor expansion of time evolution operator. If we can express an operator in the form \eqref{naveqn-4}, then it could be implemented in a quantum computer with the help of ancilla qubits. 

First, we initialize the $m$-ancilla qubit register in the following state.
\begin{equation}
	B\ket{0}=\frac{1}{\sqrt{s}}\sum_{j=0}^{m-1}\sqrt{\beta_{j}}\ket{j}
	\label{naveqn-5}
\end{equation}
where $B$ is a $m$-dimensional unitary operator and $s=\sum_{j=0}^{m-1}\beta_{j}$. We assume that there is a unitary operation select(V) that could implement each unitary $V_{j}$ and its operation can be depicted as:
\begin{equation}
	\text{select(V)}\ket{j}\ket{\psi}=\ket{j}V_{j}\ket{\psi}
	\label{naveqn-6}
\end{equation}
Where $\ket{j}$ denotes the state of ancilla qubit and \ket{\psi} will be any state. In the case of Hamiltonian simulation $\ket{\psi}$ will be initial state of the system. It turns out that the we could implement select(V) using controlled gates. To implement $\tilde{U}$, we should combine two operations \eqref{naveqn-5} and \eqref{naveqn-6}. Thus let us define another operator,
\begin{equation}
	W =(B^{\dagger}\otimes \mathcal{I})(\text{select(V)})(B \otimes \mathcal{I})
\end{equation}
then, 
\begin{equation}
	\label{naveqn-7}
	W\ket{0}\ket{\psi}=\frac{1}{s}\ket{0}\tilde{U}\ket{\psi}+\sqrt{1-\frac{1}{s^{2}}}\ket{0}\tilde{U}\ket{\Phi}
\end{equation}
for some $\Phi$ whose ancillary state is in the subspace orthogonal to $\ket{0}$. The expression \eqref{naveqn-7} gives the probabilistic implementation of $\tilde{U}$ with $1/s$ probability. In other words, by applying a projection operator $P =\ket{0}\bra{0}\otimes \mathcal{I}$ then output state will be:
\begin{equation*}
	PW\ket{0}\ket{\psi}=\frac{1}{s}\ket{0}\tilde{U}\ket{\psi}.
\end{equation*}
One can tune $s$ by varying the the size of the time interval. To convert the probabilistic implementation of $\tilde{U}$ to deterministic implementation through the method of Oblivious Amplitude Amplification (OAA). For instance, if $s=2$, one can deterministcally implement $\tilde{U}$ as given below \cite{berry15}.
\begin{equation}
	A\ket{0}\ket{\psi}=\ket{0}\tilde{U}\ket{\psi}
\end{equation}
where $A=-WRW^{\dagger}RW$. Here $R=\mathcal{I}\otimes 2P$ is called the reflection operator. The OAA is implemented by interweaving $W$ and $W^{\dagger}$ operators with the reflection operator $R$.

\subsection{Quantum Signal Processing Algorithm}

Quantum Signal Processing is a much more general approach to Hamiltonian simulation than discussed above. Although QSP takes more qubits than Trotter formulas, the method promises to scale optimally with evolution time and error tolerance. In this approach, Hamiltonian simulation is done through three steps: (i) Standard form encoding of Hamiltonian, (ii) Qubitization and (iii) Quantum signal processing. Standard form encoding embeds the hamiltonian $\hat{H}$ into the upper left block of a larger unitary matrix $U_{\hat{H}}$. Several ways are there to encode a Hamiltonian in standard form. The method to encode LCU model Hamiltonians and d-sparse Hamiltonian is well described in \cite{low19}. The next task is to perform qubitization. Qubitization is a technique for effectively mapping a multi-dimensional block encoding to a single qubit. Qubitization is performed using a pair of reflection operations. For Hamiltonians of the form \eqref{naveqn-1}, qubitization results in the implementation of the operator of the form
\begin{equation*}
	W=\exp(\pm i\cos^{-1}(\hat{H}/|\alpha|_{1}))
\end{equation*}
where $|\alpha|_{1}= \sum_{j}|\alpha_{j}|$. The next step involves the transformation of eigenvalues of $W$, $\exp(\pm i\cos^{-1}({E_{k}/|\alpha|_{1}}))$, to $\exp(-iE_{k}t)$, where $E_{k}$ is the eigenvalue of the Hamiltonian $\hat{H}$. It can be done using Quantum Signal Processing. There exist various other Quantum single value transformation methods that can perform this transformation. For instance, one can use QPE to perform this transformation if we prefer only the static quantities such as ground state energy. 
\begin{equation*}
	\exp(\pm i\cos^{-1}({E_{k}/|\alpha|_{1}})) \xrightarrow{\text{QSP or QPE}} \exp(-iE_{k}t)
\end{equation*}
The QSP, along with standard-form encoding and qubitization, implements a transformation that approximates $E_{k}\rightarrow \exp(-iE_{k}t)$ on each eigenstate of $\hat{H}$ with eigenvalue $E_{k}$, thus providing an approach to quantum simulation \cite{low19, low16}.

\section{Emerging Potential domains for Quantum Simulations}\label{5}
In this section, we discuss about the current progress towards achieving quantum advantage via quantum simulations in a variety of different areas beginning with condensed matter physics, light-matter interaction models, open quantum systems, quantum chemistry and bio-informatics.

\begin{figure*}
	\centering
	\includegraphics[width=0.5\textwidth]{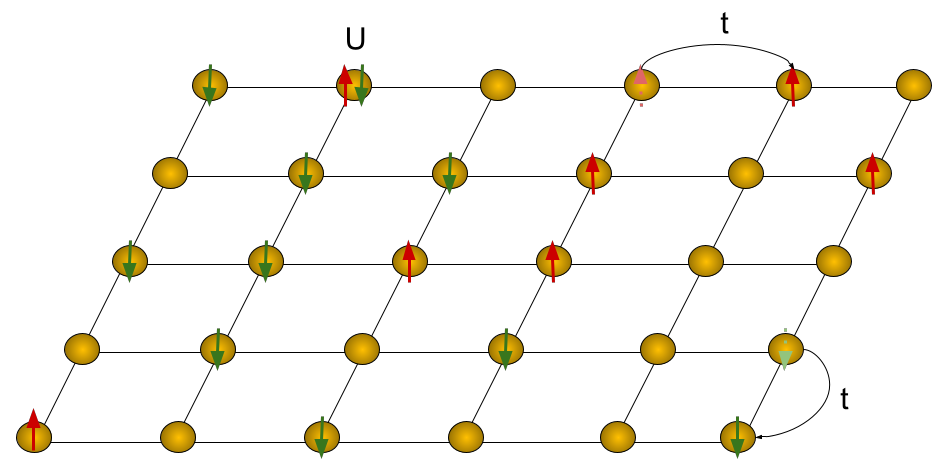}
	\caption{An illustrative representation of 2D Fermi Hubbard Model, where $U$ is the on-site interaction and $t$ is the Hopping parameter.}
	\label{Img-1}
\end{figure*}
\subsection{Condensed Matter}
The quantum material simulation is one of the paramount goals we hope to achieve through a quantum computer. Simulating exotic materials which are beyond classical computers' capabilities is crucial in developing next-generation quantum technologies. Condensed Matter Physics is the branch of science dedicated to microscopic and macroscopic properties of matter. It holds paramount importance in our understanding of nature and the development of new technologies. Although near-term quantum computers don't have efficient material simulation capabilities, numerous many-body condensed matter systems and phenomena have been simulated in them. Even though such simulations don't surpass the numerical simulations, they are considered to be baby steps toward the direction of efficient material simulation. Therefore, this section is dedicated to describing the progress of quantum simulation in the field of condensed matter physics.

In this section, we review the quantum simulation of some paradigmatic many-body systems/phenomena such as the Hubbard model, Fermion-Boson interacting systems, Spin Models, Frustrated Systems, Superconductivity and Quantum Phase Transitions. 

\subsubsection{Hubbard Model}
The Hubbard model is a quintessential fermionic system. It was proposed in 1963 by three persons independently. Hubbard put forward this model for describing the transition metals \cite{hubbard63}, and Kanamori used it to describe itinerant ferromagnetism \cite{kanamori63}, whereas Gutzwiller used it to describe the metal-insulator transitions \cite{gutzwiller63}. Technically, the Hubbard model extends the tight-binding model to include onsite interaction between particles. The tight-binding model is used to describe the motion of particles in periodic lattices but does not consider the interaction between particles within the lattice. The most commonly used Hamiltonian of the 1D Hubbard model is given by:

\begin{equation}
	\hat{H}=\sum_{\langle i,j\rangle,\sigma} t_{ij}(c_{i\sigma}^{\dagger}c_{j\sigma}+c_{i\sigma}c_{j\sigma}^{\dagger})+\sum_{i} U_{i}n_{i\uparrow}n_{j\downarrow}
	\label{naveq-1}
\end{equation}
where $t_{ij}$ is called hopping parmeter usually it is taken as constant $t$. Here $\langle i,j\rangle$ denotes nearest neighbouring sites (one can also consider next-nearest neighbour hopping and other higher orders). The $c^{\dagger}_{i} (c_{i})$ is the creation and annihilation operator of particle at the site $i$, and $n_{i\uparrow}=c^{\dagger}_{i} c_{i}$ is the number operator that counts number of particles occupying at a particular site $i$. The onsite interaction energy is represented by $U_{i}$, which is usually assumed to be a constant $U$. Originally it was proposed to study the behaviour of correlated electrons (fermions) and hence $U$ will be coulombic repulsive interaction. But in general, the nature of $U$ can be either attractive or repulsive. Hubbard Model can also modified to describe the physics of the spinless Bosons interacting in the lattice. In the former case, the model is called the Fermi-Hubbard Model (F-H Model), and in the latter, it is called Bose-Hubbard Model (B-H Model) \cite{fisher89}. The Bose-Hubbard Hamiltonian is given by:
\begin{equation}
	\hat{H}=t\sum_{\langle i,j\rangle,\sigma} b_{i\sigma}^{\dagger}b_{j\sigma}+\dfrac{U}{2}\sum_{i} n_{i}(n_{i}-1)
	\label{naveq-2}
\end{equation}
In the Bose-Hubbard Hamiltonian, the fermionic operators are replaced by corresponding bosonic operators since it describes Bosons. Also, in some descriptions of the Hubbard Model, there would be an additional term in the Hamiltonian \eqref{naveq-1} \& \eqref{naveq-2}, corresponding to the chemical potential, which sets the total number of particles. 

The F-H model is a radically simplified version of the much more general Electronic structure Hamiltonian, which has applications in quantum chemistry \cite{wecker15}. The Physics of the Hubbard model is governed by the $U/t$ ratio (potential to kinetic energy ratio). Consider a system modelled by a repulsive F-H model of spin 1/2 particles with each site accomodating utmost two particles (two-single particle states). It would have a metallic nature when $U/t=0$, since only hopping occurs. For $U/t << 1$, the system will have an Antiferromagnetic phase (spins in a site will be anti-aligned due to Pauli's Exclusion Principle). The antiferromagnetic Heisenberg model Hamiltonian can be derived from the Hubbard Hamiltonian \cite{wecker15} under certain assumptions. The system will have a band insulating nature when all sites are completely filled (n=2) for  $U/t << 1$. Now when $U\approx t$, under certain conditions, the system will have a paramagnetic phase \cite{spalek2007}.
\end{multicols}
\begin{figure}[h]
\centering
\includegraphics[width=0.8\textwidth]{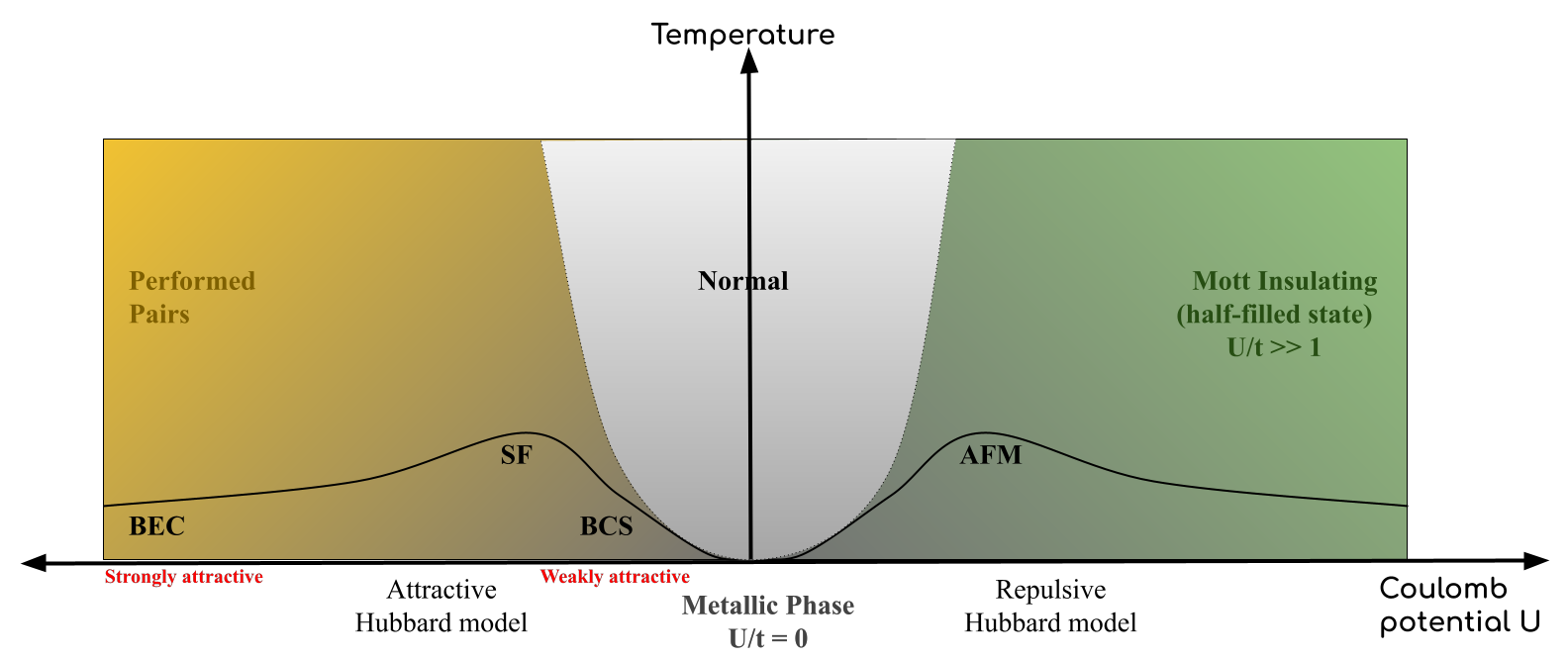}
\caption{A representational phase diagram for the attractive and repulsive Hubbard model at half filling \cite{esslinger10}. Abbrevations: AFM- Anti-ferromagnetic phase,  BCS- Bardeen-Cooper-Schrieffer (Superconducting Phase); SF- Superfluid, BEC- Bose-Einstein condensation.}
\end{figure}
\begin{multicols}{2}
Further, if each site is occupied by one particle (n=1 ``half-filled state"), then for $U/t >>1$ system will have a Mott-Insulating Phase. Mott Insulators are essentially different from band insulators. Even though sites are half-filled, hopping cannot take place due to high coulombic repulsion. Thus it will have an insulating nature rather than a conducting one. The Fermi-Hubbard model captures the metal-insulator (``Mott-Hubbard") phase transitions. The system will have a metallic phase when $U<<t$ and a mott-insulating phase at $U>>t$. Mott Insulators are essentially different from band insulators. Even though sites are half-filled, hopping cannot occur due to high coulombic repulsion. Thus it will have an insulating nature rather than a conducting one. Hence the system will have a metallic phase when $U<<t$ and a mott-insulating phase at $U>>t$ and the F-H model captures the metal-insulator (``Mott-Hubbard") phase transitions.

The attractive F-H model is extensively studied in the context of superconductivity \cite{micnas90}. The BCS (Bardeen-Cooper-Schrieffer) regime is characterised by weak attractive interaction and strong-attractive interactions can result in bound pairs that could undergo Bose-Einstein Condensation (BEC)\cite{nozires85}. At very low temperature, BEC-BCS crossover is expected. In addition, the t-J model, which can be used to study High-Temperature superconductivity in cuprates, can be derived from the Hubbard Model. In comparison with Mott-insulating regime of Hubbard model  t-J model is characterised by $U/t >>1$, and $n \neq 1$ (i.e. system is doped).

The Bose-Hubbard model came into importance with the development of trapping Ultra-cold atoms in optical lattices. It was the first strongly correlated model that was realised using ultracold atoms in optical lattices. Ultra-cold atom experiments with bosonic were able to observe the Superfluid to Mott-insulating phase transitions \cite{greiner2002}. Moreover, the Bose-Hubbard model can provide compelling descriptions of systems such as dilute alkali gases in optical lattices and arrays of Josephson junctions.

The exact solution of the 1D F-H model was given by Lieb and Wu in 1968 using Bethe Ansatz \cite{lieb68,lieb2003}. But at Higher dimensions, the Hubbard Model is not exactly solvable. Various numerical treatments are there to study such models. Approximate methods based on perturbative theory fail in highly interacting or strongly correlated regimes. It is because of the non-perturbative nature of the problem. In such scenarios, one can numerically perform exact diagonalisation of finite size systems or use Quantum Monte Carlo (QMC) methods. But exact diagonalisation requires computational resources that would grow exponentially with system size. Thus only applicable for systems of small size. Now, QMC methods allow the evaluation of phase-space integrals through random sampling of phase space. But for strongly correlated fermionic systems, QMC methods suffer from ``sign problem" at low temperatures. It arises due to the difficulty of numerically evaluating the integral of an oscillatory function. During integration, the positive and negative contributions of integrals nearly cancel out. The sign problem can be circumvented using Determinant Diagrammatic Monte Carlo methods, which perform stochastic summation of Feynman diagrams with controllable error bars \cite{boninsegni2006,van2010}. But even then these approaches suffer from significant finite-size effects, making it challenging to extract low energy scales, which are critical for capturing the competition between distinct ground states common in strongly correlated systems.  

Mean-field theories are immune to finite-size problems and often act as a way to study these models. Dynamical Mean Field Theory (DMFT) is one of the standard approaches used to study correlated systems. DMFT maps the lattice model into the Anderson Impurity model, which is solvable through various methods. DMFT allows one to calculate the local lattice greens function. One can be used to calculate several physical properties like the spectral function (which gives the band structure), the kinetic energy, the double occupancy of a site and response functions (compressibility, optical conductivity, specific heat) from the local lattice greens function. Another method used to extract the dynamical and static properties of the system is the Lanczos Algorithm. It is an iterative algorithm that could be used to find the eigenvalues and the eigenvectors of Hermitian matrix.

\subsubsection{Quantum simulations of Hubbard Model}\label{Quantum simulation Hubbard Mode}
Analog simulators, particularly the ultra-cold atom simulators, are among the favourite architectures for simulating the Hubbard model and the fermionic and bosonic systems. The high degree of controllability and the observation tools based on the Quantum Gas Microscope enables individual atom-level detection. It allows one to measure intricate correlations between constituent particles that are difficult to calculate using other architectures. In addition, the Hubbard model could easily be realised using ultra-cold atoms in the optical lattices. Some of the exciting works and results in this area include the observation of superfluid-mott insulating phase transition in the B-H Model \cite{greiner2002}, observation of mott-insulating phase in F-H model \cite{jorden2008}, Measurement of the equation of state in the density sector of two dimensional Hubbard model \cite{cocchi16}, Measurement of nearest neighbour correlation in lower and higher dimensions \cite{greif13,greif15,hart15} and direct detection of spin-charge multipoint correlations \cite{hilker17}. In addition, dynamical non-equilibrium simulations could be also performed \cite{schneider12,greiner2002b,chen11,schafer20} which is hard to perform even on supercomputers.

In the digital quantum simulation (DQS) fermionic systems, one could use either the first or second quantized description of Hamiltonian. But often, the DQS using the first quantized description would require more resources in terms of qubits and gates than using the second quantized description \cite{abrams97}. Thus, in most DQS of F-H models, we use second quantized formalism for simulation. Nevertheless, using the first quantized description will be advantageous in cases where the number of fermions is significantly less than the number of single-particle states in the system. To perform DQS, various quantum computing platforms like superconductor based, ion trap based and cold-atom based are available. But because of its ability to implement long-range interactions easily, ion trap based simulations of Fermionic and Bosonic systems are considered efficient compared to the others \cite{lamata14}. In addition, the ability to implement the non-local Molmer Sorenson gate directly ease the simulation of dynamics of fermion-boson interaction. Further optimization schemes exist to make the entangling gates faster and reduce the simulation time.

The DQS of the F-H model or fermionic systems, in general, involves three steps: (i) Mapping the system to the qubit, (ii) Performing Hamiltonian evolution, and (iii) Measuring observable. While simulating the F-H model, Jordan-Wigner transformations are generally used for mapping and Hamiltonian evolution performed through trotterization (Product formula). The simulation of the 1D F-H model with nearest neighbour interactions is discussed in \cite{barends15,heras15}. In both works, they have studied the state evolution with time. In addition, \cite{barends15} have simulated the effects of a time-varying interaction in the system. Comparing AQS, there is little resource in the simulation of the Hubbard model (fermionic systems) in higher dimensions using DQS. To simulate the 2D F-H model using superconducting architecture was put forward in cQED architecture \cite{heras15} where pulses could be used to implement various interactions. Their design could efficiently implement nearest and next to nearest interactions. In gate-based quantum computing, there exist a mapping scheme which could map $N_{x}\times N_{y}$ 2D lattice in to 1D chain of $2(N_{x}\times N_{y})$ qubits \cite{somma2002}. One could use this mapping scheme to simulate the 2D Fermi Hubbard model through DQS. 

Digital Quantum Simulations takes a universal approach and thus could be used to evaluate various properties that are not accessible through experiments. For example, in \cite{linke18} they have demonstrated the calculation of Renyi entropy on a two-site F-H model. Additionally, the non-equilibrium dynamics of the F-H Hubbard model can also be studied using DQS. A recent work discussed the time and momentum distribution of the fermionic distribution function and the evolution of entanglement entropy due to the quenching of interaction potential of the F-H model \cite{fauseweh21}. Similarly, separation of spreading velocities of charge and spin densities after quenching onsite interaction and trapping potential in the F-H model was also recently observed \cite{arute20}. Another approach to studying non-equilibrium dynamics is using an open quantum system. In this approach, the open quantum system is explicitly simulated. Using this approach, a dissipative two-site Hubbard model is simulated lately \cite{tornow21}. 
\subsubsection{Fermion-Boson Interaction systems}
There are different physical systems where fermion-boson interactions play a significant role in governing the dynamics. In condensed matter theory, studying fermion-boson interactions is crucial to modelling the electron-phonon interaction in solids. Especially in metals, the Electron-Phonon Coupling (EPC) would influence the low energy electronic excitations, influencing the thermodynamic and transport properties in metals. Furthermore, charge density wave (CDW) order in transition metal dichalcogenides and high-temperature superconductivity in Bismuthates are both explained by electron-phonon interactions. Generally, we use three models to study the EPC: (i) The Froehlich Model, (ii) The Holstein Model and, (iii) Su-Schrieffer-Heeger (SSH) Model or Peierls Model. 

In the Froehlich model, unscreened coulombic interactions couple the electrons o the longitudinal acoustic modes of phonons. Here, we do not assume any lattice thus will be a jellium model. The Holstein model is inherently a lattice model-based description of the EPC. It is similar to Hubbard's model used to describe strongly correlated fermions. Holstein's model assumes the electron coupling with the single branch of the optical phonon. 

Polarons are quasiparticles formed as a result of electron-phonon interactions. Polarons were originally proposed to describe an electron moving in a dielectric crystal. In such systems, displacement of atoms from their equilibrium positions effectively screens the charge of an electron, known as a phonon cloud. Generally, Polarons are described using either Froehlich Hamiltonian or Holstein Hamiltonian. When a polaron's spatial extension is greater than the solid's lattice parameter (polarizable continuum), it is called the Froehlich polaron. A Holstein polaron can form when the self-induced polarization created by an electron or hole is of the order of the lattice parameter. The production of polarons can significantly reduce electron mobility in semiconductors. Organic semiconductors are also susceptible to polaronic effects, vital in constructing effective charge-transporting organic solar cells. Polarons are also helpful in deciphering the optical conductivity of these materials. There are also significant extensions of Polaron concepts that could investigate the characteristics of conjugated polymers, high Tc superconductors, layered MgB2 superconductors, fullerenes quasi-1D conductors, and semiconductor nanostructures.

Apart from the two models, the SSH model takes a different approach to describe the coupling. Rather than through potential energy, electrons and phonons are coupled through kinetic energy. As a result, the EPC shows electron momentum dependence in the SSH Model. This model describes spinless fermion hopping in a one-dimensional lattice with a staggered hopping parameter. It was originally developed to model polymers like polyacetylene. However, it also models materials with strong electron-phonon couplings, such as nanotubes, C60 molecules, and other fullerenes.

Numerous ways are there to solve Polaron models and extract the properties. The first approach is the exact diagonalization of the model using computational resources. Even though it works for smaller systems \cite{alexandrov94,marsiglio95,barisic2006}, it becomes intractable with an increase in system size. Thus there are other approaches, such as one based on variational methods \cite{romero99,barisic2004,bonca99}, and one based on Quantum Monte Carlo methods \cite{cohen22,kornilovitch98}. A review of variational and QMC methods for Holstein models is given in \cite{hohenadler2004}. One could also apply Diagrammatic Monte Carlo methods to calculate the Greens function of such models \cite{macridin2003}. Finally, approaches such as the Density-Matrix Renormalization Group for one-dimensional systems \cite{jeckelmann98} and dynamic mean-field theory (DMFT) for infinite-dimensional systems apply to certain circumstances \cite{ciuchi97}.

\subsubsection{Quantum simulation of fermion-boson interactions}
There are very few resources in this field, and most of the work in this direction is done in the trapped ion architecture. In trapped ion architecture, the bosonic systems can be mapped to the vibrational states of ions. Possession of additional degrees of freedom in addition to qubits makes trapped ions promising for implementing fermion-boson interactions. Further, the ability to implement non-local gates such as the Molemer Sorenson gate directly in trapped ions significantly reduces the gate counts while simulating the fermion-boson interacting system \cite{lamata14} than other architectures. It has been shown that a many-body fermionic lattice model can be simulated on an ion-string \cite{casanova12}. Two trapped ions\cite{casanova11} have been used to simulate Fermion and antifermion modes in a bosonic field mode. It is one of the simplest systems in Quantum Field Theory where fermion-boson interaction occurs.

In the case of DQS, there is a lack of algorithms that would efficiently simulate the evolution of boson states. It makes it hard to simulate fermionic-boson coupling in quantum computers. Nevertheless, recently a quantum algorithm was put forward to simulate fermion-boson interaction \cite{macridin18}. This algorithm maps bosons to qubits after treating bosonic degrees of freedom as finite-set of harmonic oscillators. They also benchmarked this algorithm by simulating a 2-site Holstein polaron model. Digital quantum simulation of the Holstein model in trapped-ion has been discussed in \cite{mezzacapo12}.
\subsubsection{Spin model}
Spin Models are one of the simple models used to describe various phenomena. But it is often not used to provide an accurate rather qualitative description of the system. However, they provide a quantitative description of the properties of mott insulators with localized electron spin. Even though they are primarily used to model magnetism, their uses are far-reaching such as solving optimization problems and restoring digital images. Besides that, spin models are extensively used in Quantum Field Theory (Integrable model), Quantum Information Theory (for testing many-body concepts like Entanglement entropy) and Quantum computation. 

Primarily, spin models are classified into two: Classical Spin Models and Quantum Spin Models. Both models' theoretical formulation is fundamentally different. One of the striking differences is the non-commutability of spin observables in quantum spin models compared to classical spin models. Also, both models serve different kinds of use. The classical spin models have been crucial in studying thermal phase transitions and critical phenomena. In contrast, quantum spin models provide a theoretical framework for quantum magnetism and quantum phase transitions. In addition, effects like fractional quantum effect, frustration, disorders, and so on open a new realm of exotic many-body states described using quantum spin models. Examples of classical spin models include classical Ising chains, classical Heisenberg model and Potts model (model for interacting spins in the crystal lattice). Quantum Ising Model (or Transverse Field Ising Model), Quantum Heisenberg Model, and Kitaev Model (neighbouring spins interact with anisotropic Ising interactions)\cite{hermanns18}. We are only interested in the quantum spin model and focus discussions in that direction. In the following paragraphs, we start a discussion with the quantum Heisenberg model and describe its relation with other models. The Heisenberg Hamiltonian of a system of $N$-spin with nearest-neighbour interaction is given by:
\begin{equation}
	\hat{H}= J_{x}\sum_{<i,j>} \sigma^{x}_{i} \sigma^{x}_{j} +J_{y}\sum_{<i,j>} \sigma^{y}_{i} \sigma^{y}_{j}+J_{z}\sum_{<i,j>} \sigma^{z}_{i} \sigma^{z}_{j}+h\sum_{i}\sigma^{z}_{i}
	\label{naveq-3}
\end{equation}
In the description of Heisenberg Hamiltonian \eqref{naveq-3}, we implicitly assume that the exchange interactions $J_{x}$,$J_{y}$ and $J_{z}$ to be the same for all spins and the magnetic field to be along the $z$-direction. Here $h$ represents the strength of the magnetic field. Various spin models could be derived from the Heisenberg model. If $J_{x}\neq J_{y}\neq J_{z}$ it is called XYZ Heisenberg Model. If $J_{x}=J_{y}\neq J_{z}=\Delta$ it is called XXZ Heisenberg Model. If $J_{x}=J_{y}=J_{z}$, it is called XXX Heisenberg model. If $J_{z},h=0$ it is called XY model and additionally if $J_{x}=J_{y}$ it is called XX model. One can construct disordered spin chains from these models by uniform random sampling of $h$ within an interval.  One of the simplest extensions of the Heisenberg model is the $J_{1}-J_{2}$ model. In addition to nearest neighbour interaction, it also includes next to nearest neighbour interaction.

The transverse field Ising model could be derived from \eqref{naveq-3} by taking $J_{y}=J_{z}=0$. To model Ferromagnetism and anti-ferromagnetism, we usually take the exchange coupling parameters as a constant value $J$ (also $h=0$, i.e. no external field). If $J<0$, then neighbouring spins align parallelly to minimize energy and thus represents a ferromagnetic system. In contrast, if $J>0$, neighbouring spins anti-align to reduce energy and thus represents an anti-ferromagnetic system.

Kitaev model also could be related to Heisenberg Model. Kitaev model is described on a honeycomb lattice composed of two sub-lattices. It has anisotropic nearest neighbour spin interaction that depends on the direction of the bond and is given by:
\begin{equation}
	\hat{H}=-\sum_{x-link}J_{x}\sigma^{x}_{i}\sigma^{x}_{j}-\sum_{y-link}J_{y}\sigma^{y}_{i}\sigma^{y}_{j}-\sum_{z-link}J_{z}\sigma^{z}_{i}\sigma^{z}_{j}
\end{equation}
The x, y, and z links in the hexagonal lattice are three separate bonds connected by a 120-degree rotation, and the $i,j$ represents the nearest neighbour sites. The model is thought to help explain strongly correlated materials. It has been presented as a model for a variety of condensed matter systems, including quantum dots coupled to topological superconducting wires, graphene flake with an irregular boundary, and kagome optical lattice with impurities. Spin lattice models with bond-dependent Heisenberg coupling are frequently used to study the exotic Quantum Spin Liquid phase. In honeycomb lattice, QSL is modelled using the Kitaev model.

Both the Heisenberg model \cite{karbach98,karbach98b}and Kitaev model \cite{mandal20} is exactly solvable in 1D. Interestingly there exist an encyclopedia of exactly solvable many-body systems in 1D \cite{mattis93}. Now in the case of 2D, under certain assumptions, the Heisenberg model could be solved \cite{sriram81,affleck88} but no general solution exist. Thus in Higher dimensions, we resort to numerical simulation. Numerical exact diagonalization and Quantum Montecarlo methods used in quantum spin systems are reviewed in \cite{sandvik10,todo01}. In addition, the Density Matrix Renormalization Group (DMRG) approaches are also found in quantum spin systems \cite{white93, wiki,yan11,jiang12}. 
\subsubsection{Quantum Simulation of Spin Models}
Various quantum simulating architectures are used to simulate Spin models. For instance, using trapped ions, simulation of Ising, XY, XXZ spin chains in \cite{porras04}. In their scheme, they couple internal states with vibrational modes for simulation, which enable them to observe rich phase transition in ion traps. Simulation of transverse field Ising model is also discussed using  $^{171}Yb+$ ions \cite{edwards10,kim11}. Similarly, the adiabatic evolution of TFIM from paramagnetic to ferromagnetic order is discussed in \cite{friedenauer2008}. Moreover, the emergence of magnetism is displayed by implementing a non-uniform ferromagnetic quantum Ising chain using nine trapped $^{171}Yb+$ ions. In addition, a frustrating system has been simulated using the Ising type of interaction between three trapped ions \cite{kim10}. There have also been attempts to move out the nearest neighbour interactions and include higher range interactions in trapped ions. Triangular 2D lattices made using Beryllium ions stored in Penning traps have been used to variable range Ising interactions between spins \cite{britton12}. There are even discussions about simulating nonlinear spins models in trapped ions \cite{milburn99}.

There are works in Spin model simulations on other architectures also. There have been investigations into simulating spin chains and ladders using atoms in optical lattices \cite{garcia04,simon11}. Similarly, Itinerant magnetism has been studied using two-component fermi gas \cite{jo2009}. There also has been a discussion on simulating the spin-lattice model using \cite{micheli2006}. There also had been some exciting works in cQED architectures on spin models. A scheme for simulating an anisotropic XXZ chain using coupled cavities is described in \cite{cho2008}. In \cite{tsomokos10}, simulation of spin models in higher and fractal dimensions has been proposed. Since superconducting qubits can easily access more than two states, there also have been discussions to simulate spin models with different spin quantum numbers \cite{neeley2009,nori2008}. In addition, using superconducting circuits, Floquet quantum simulation of spin models has been carried out \cite{kyriienko18}.
\begin{center}
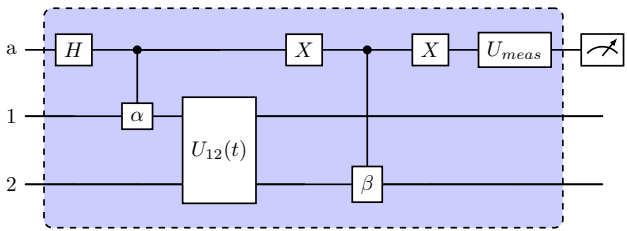

	\begin{tikzpicture}
		\node[scale=0.8]{
			\begin{quantikz}
				\lstick{a} & \gate{H} \gategroup[3,steps=7,style={dashed, rounded corners,fill=blue!20, inner xsep=2pt, inner ysep=8pt}, background]{{\sc}} & \ctrl{1} & \qw & \gate{X} & \ctrl{2} & \gate{X} & \gate{U_{meas}} &\meter{}\\
				
				\lstick{1} & \qw & \gate{\alpha} & \gate[wires=2] {U_{12}(t)} & \qw & \qw &\qw & \qw & \qw\\
				
				\lstick{2} & \qw & \qw & \qw & \qw  & \gate{\beta} & \qw &\qw & \qw
			\end{quantikz}
		};
	\end{tikzpicture}
	
	\captionof{figure}{Ancilla based algorithm for measuring dynamical cross-coorelations in spin dimers. $U_{12}(t)$ simulate the interaction of spins with external field. The gates $\alpha$ and $\beta$ represents the Pauli gates. $U_{meas}$ permits retrieving the expectation values of $\sigma_{x}$ (for $U_{meas} = H)$ and $\sigma_{y}$ (for $U_{meas} = R_{x}(\pi/2)$) on the ancilla (``a") \cite{tacchino}.}
\end{center}
Most quantum computing architectures can directly map Spin states to the qubit states. Thus simulating spin models through DQS is direct and does not require any transformation schemes. The TFIM is one of the simplest models that could be simulated in DQS through trotterization \cite{powers21}. The Hamiltonian is so simple that it could be exactly simulated \cite{cervera18}. In the exact simulation, one has to construct the disentangling operator that would transform Hamiltonian to be diagonal on the computational basis. In the case of the Ising model, using Jordan Wigner, Fourier and Bouglibov transformation in succession, one could realize the disentangling operator to perform exact simulation. The DQS Heisenberg model and its variants are discussed in-depth in \cite{tacchino}. They have also used an ancilla based algorithm to extract the spin-spin dynamical correlations of the models discussed. Similarly, simulation of the Heisenberg model through DQS is also addressed in cQED architecture \cite{heras14,salathe15}, which could be easily translated to gate based algorithm. In the case of the honeycomb Kitaev model, very recently, a protocol has been proposed to simulate its ground state \cite{bespalova21}. Moreover, the far-from-equilibrium dynamics of spin chains (including disordered chains) are simulated in \cite{smith19}. To probe the far-from-equilibrium dynamics, they performed quantum quenching and studied it by measuring various observables like Magnetization, Connected equal-time correlator and Quantum Fisher Information.

There also had been discussions in \cite{jane2002} on realizing universal quantum simulation using quantum optical elements. They have discussed methods of simulating Ising and Heisenberg Hamiltonian in such simulators. Similarly,  universal digital quantum simulation of the Ising and Heisenberg model with trapped ion is discussed in \cite{lanyon11}. Further, there are some works in the digital-analog simulation of spin models, as well \cite{gonzalez21,arrazola16}.

\subsubsection{Frustrated systems}
Strongly correlated systems are prone to frustration when the system cannot meet all the constraints imposed by Hamiltonian simultaneously. One paradigmatic example is the system having triangular lattice and anti-ferromagnetic (AF) ordering. The third spin will be frustrated once the other two spins align anti-parallel due to AF order  \cite{wannier50,moessner2001}. It means the third spin cannot minimize interactions with the others simultaneously. A similar type of frustration can also be seen in the Kagome lattice (a 2D lattice structure found in many natural minerals) with AF order. But frustration is not limited to systems with AF ordering. It can also happen in systems with competing interactions. For example, the system will be frustrated if one replaces an odd number of ferromagnetic bonds with anti-ferromagnetic ones in a ferromagnetic Ising model. Similarly, the square lattice with ferromagnetic ordering, except for one site with anti-ferromagnetic interaction, shows frustration due to competing interactions. Till now, we discussed frustration happening in regular lattices. It is called Geometrical Frustration. But frustration can also occur in disordered systems, such as spin glasses and spatially modulated magnetic superstructures. In such systems, frustration will be amplified by disorderliness. 
\end{multicols}
\begin{figure}
\centering
\includegraphics[width=0.8\textwidth]{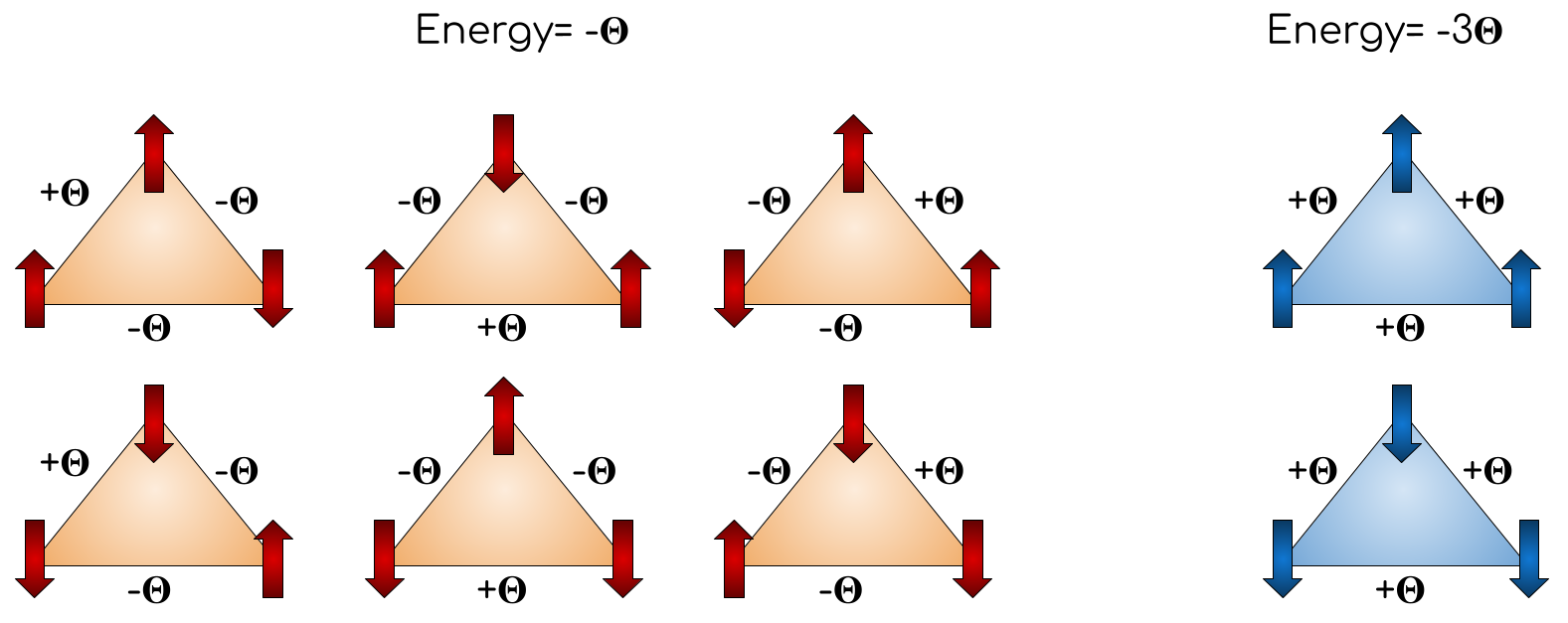}
\caption{At zero temperature and magnetic field, all conceivable configurations of a three-spin frustrated Ising magnet \cite{zhang12}. The ground state is sixfold degenerate, resulting in a non-zero entropy. Here $\Theta$ denotes the external magnetic field.}
\label{Img-3}
\end{figure}
\begin{multicols}{2}
Frustrated systems show ground state degeneracy. For example, the ground state of the triangular lattice with AF order is six-fold degenerate. These systems also have rich Phase diagrams, and fascinating effects occur when the system is both disordered and frustrating \cite{sachdev99}. Further, unlike ordinary systems, a frustrated system shows non-zero entropy when the temperature tends to absolute zero.

Numerical simulations are used to study frustrated systems. Exact diagonalization is one of the approaches used. There are numerous exact diagonalization algorithms such as the Lanczos algorithm \cite{lanczos50}, Davidson algorithm \cite{davidson75} and, Lapack/Householder complete diagonalization method \cite{meisner2006} with fast convergence. But the downside of the exact diagonalization method is the exponential scaling of resources with the system size. Other approaches include Classical Monte Carlo methods\cite{weber2003,alet2005,alet2006,huse2003}, Quantum Monte Carlo methods \cite{wessel2005,melko2005}, Density Matrix Renormalization Group methods \cite{chitra95,white2007,jiang2008} and, Series expansion methods \cite{singh2007,rigol2006,trebst2000}.
\subsubsection{Quantum Simulation of Frustrated Systems}
Quantum simulations provide an alternative to studying frustrated systems. It is believed to be more advantageous than classical methods. Recently, in the quantum simulation of geometrically frustrated magnets, a scaling advantage over path-integral Monte Carlo method has been demonstrated utilizing a superconducting flux-qubit quantum annealing processor \cite{king21}. Moreover, numerical methods often have a hard time simulating the ground state of frustrated systems, which is highly entangled. In AQS, the ground state of the frustrated Ising spin has been simulated, and its properties have been studied in a system of three trapped ions \cite{kim2010}. They have performed adiabatically evolution of the state from a transversely polarised state.  Photonic quantum simulators are also considered to be advantageous in simulating correlated chemical or solid systems since entangled photon states could be easily generated in it \cite{guzik2012}. Photonic quantum simulators have simulated the ground state wavefunction of four spin-$1/2$ systems with Heisenberg interactions \cite{ma11}. In addition, measurement induced interactions in a four spin-system, simulated using a photonic simulator, were carried out in \cite{ma15}. Further, the spin correlations and coherent dynamics of frustrated systems were investigated utilizing spins held in a crystal with up to 16 trapped $^{171}Yb+$ atoms. Besides, there are discussions about the simulation of frustrated systems in Ultracold atom \cite{struck2011} and NV centre-based \cite{cai13} quantum simulations.

In DQS, works in this direction are scarce. There has been a work in NMR quantum computer on thermal (Gibbs) state simulation of a frustrated system through the Coherent Encoding of Thermal State (CETS) \cite{zhang12}. The method can be extended to all gate-based quantum computers. Once the Gibbs state is simulated, all the relevant thermodynamical properties of a system can be extracted. Similarly, the simulation of frustrated anti-ferromagnetic Heisenberg chain has been performed in cQED architecture, which could be easily translated into gate based simulation \cite{heras14}.

\subsubsection{Superconductivity}
Superconductors are materials that can carry electric charge without resistance while also exhibiting macroscopic quantum phenomena, including persistent electrical currents and magnetic flux quantization. Examples of such materials include Aluminium, Niobium and Gallium so on. Unlike ordinary electrical conductors, whose resistance only becomes zero at absolute zero, superconductors have zero resistance below a critical temperature $T_{c}$. In addition, these materials display intriguing properties such as the Meissner Effect, Persistent Currents and Critical Currents. One of the first approaches to explain superconductivity was the Bardeen-Cooper-Schrieffer's (BCS) theory. According to BCS theory, the electrons inside the material condense into cooper pairs below the critical temperature. A cooper pair is a bosonic particle formed through the pairing of two electrons via a phonon. Thus at very low temperatures, the lattice vibrations will not be strong to break the cooper pairs, and hence there will not be resistance. Now, the BCS hamiltonian belongs to the class of Pairing Hamiltonian, which is used to describe paring interactions in nuclear physics and mesoscopic condensed physics \cite{moller92, braun99}.

The presence of an energy gap $\Delta$ between the BCS ground state and the first excited state is a critical aspect of BCS theory. It's the least energy necessary to excite a single electron (hole) from a superconducting ground state. Thus a Cooper pair's binding energy is two times the energy gap $\Delta$. The energy gap is temperature-dependent. The energy gap will be zero at critical temperature ($\Delta (T_{c})=0$) and will be approximately equal to $1.76K_{B}T_{c}$ at absolute zero (($\Delta (0)=1.76K_{B}T_{c}$). Finding the energy gap of material is important in BCS theory as it manifests in various thermodynamical properties of the material, such as density of states, free energy, low-temperature specific heat and so on \cite{fernandes20}.

Now, the BCS theory is not complete. There are superconducting materials which do not obey BCS theory. Those explained by BCS theory are called conventional superconductors, and most metallic superconductors belong to this category. But there are unconventional superconductors characterized by a critical temperature greater than 77 K, the boiling point of liquid Nitrogen (cheaper than liquid Helium). Most high-$T_{c}$ superconductors are ceramic materials that are not conductors at room temperature. Examples of high-$T_{c}$ superconductors include Cuprates, Magnesium dibromide, Nickelates and so on. High-$T_{c}$ superconductivity is a phenomenon that we know little about and is an active area of research. The high-$T_{c}$ superconductivity is modelled using the $t-J$ model. It could be derived from the repulsive Hubbard model at the $U>>t$ limit. Since onsite repulsion is significant, it will not allow two electrons to occupy a site and would have a conducting nature \cite{anderson13,dagotto94,spalek2007}.

The pairing Hamiltonians are exactly solvable \cite{richardson64}. Further, BCS hamiltonian is also integrable and solvable using Bethe Ansatz. But $t-J$ model is not solvable even in 1D. However, using Bethe Ansatz, one could solve the $t-J$ model for specific ratios of $J$ and $t$ in 1D \cite{sarkar90}. In general, numerical methods are used to solve such systems. The exact diagonalization is a popular approach for analyzing the $t-J$ model quantitatively. This approach employs the Lanczos algorithm, which yields accurate results \cite{lee2004,hasegawa89,ogata91}. Lanczos algorithm could also be used to calculate the ground-state properties \cite{corboz14,hellberg99}. Variational Monte Carlo is another approach to simulate the $t-J$ model \cite{corboz14,morita15}. 
\subsubsection{Quantum Simulation of Superconductivity}

The AQS of the $t-J$ model is proposed in \cite{yamaguchi2002} using a non-doped parent crystal $La_{2}CuO_{4}$ as a simulator. The high-temperature superconductivity of compounds with copper-oxide planes is still a mystery that large-scale simulations might help answer  As indicated in \cite{manousakis2002}, the $CuO_{2}$ plane in a high-Tc superconductor might be analogously simulated by an array of electrostatically defined quantum dots. 

In DQS, a polynomial-time algorithm for simulating Pairing Hamiltonians using NMR quantum computers has been put forward in \cite{wu2002, brown2006}. The energy gap of the system is found through simulation with polynomial resources. A modified version of this algorithm applicable to Qubus ancilla driven quantum computation is described in \cite{brown11}, which could be translated to gate based simulations. Recently, a hybrid quantum simulati on approach was put forward to obtain the energy gap from the BCS Hamiltonian \cite{sa22}. They use the Variational Deflation Algorithm that gives the energy of the excited state to obtain the energy gap. In addition, discussions of solving the BCS gap equation are there in \cite{bassman21}.

\subsubsection{Quantum phase transitions}
Understanding phase transition is one of the main objectives of science. Phase transitions (PT) are part of everyday life, and the paradigmatic transitions are those between Ice, Water and, Vapour. Even in the formation of galaxies, stars etc., PT play a significant role. Phase transitions, by and large, are of two classes: (i) Classical Phase Transition (CPT) and (ii) Quantum Phase Transition (QPT). In the following paragraphs, we will broadly discuss about both.

Classical Phase Transitions (CPT) are those we daily encounter in life. Thermal fluctuations drive these sorts of transitions. Depending on the discontinuity of the derivative of a thermodynamic potential, PTs are further divided into First-order and Second-order (Continuous) Phase transitions. The boiling of water at $100^\circ C$ is the first-order PT, whereas the ferromagnetic transition of iron above  $770^\circ C$ is a second-order PT. In both PTs, there would be a quantity called order parameter, which takes zero value in the disordered phase and non-zero value in the ordered phase. In the first case, it will be the difference in the densities of liquid and gas phases, and in the latter, it will be the Magnetization. Classical Phase Transition is also called Thermal Phase Transition.

Quantum Phase Transitions (QPT) are fundamental of a different origin than the CPT. Rather than thermal fluctuations, Quantum fluctuations drive a QPT. Since thermal fluctuations dominate quantum fluctuations at non-zero temperatures, a QPT only occurs at or near absolute zero. Hence a QPT is due to variation of non-thermal parameters in the Hamiltonian such as pressure, magnetic field or chemical composition. Most of the QPTs are second-order, but they can also be of the first order. One example is the topological fermion condensation QPT (in strongly correlated quantum spin liquid). A detailed description of QPT is given in \cite{sachdev99,vojta2003}.

We will discuss some interesting Quantum Phase transitions in the following few paragraphs. One of the commonly studied QPTs is the Metal-Insulator transition. It happens in a variety of systems where one can tune interatomic distances. Some examples are the PTs found in atomic or molecular gases at low temperature and high pressure and the PTs occurring in disordered systems when the degree of disorder or electron density is changed. Likewise, the repulsive Hubbard model exhibits a metal-insulator at low temperature by tuning coulombic interaction. Even topological insulators also exhibit similar phase transitions. In the presence of parallel magnetic, ultra-thin films of topological insulators show semi-metallic to an insulator PT. Due to recent advancements in fields like many-body localization, fractional and quantum hall effects and high-temperature superconductivity, QPT in electronic systems \cite{vojta2000} has gained more attention.

Another intriguing class of QPT is the Superconductor-Insulator PT. Bose Hubbard model exhibits interaction or disorder driven Superfluid-Mott insulating PT. Charged Superfluid (i.e., superconductor) undergo a PT to the insulating phase by varying magnetic field. In addition, disordered thin films also show this transition. Moreover, Cuprate Superconductors, a quintessential high-temperature superconductor, exhibit Mott insulating to a d-wave superconducting phase transition while doping. 

Magnetic systems also display Quantum Phase Transition. A typical example is a ferromagnetic-to-paramagnetic PT in Transverse Field Ising Model (TFIM). Most of these types of transitions are modelled using simple Quantum Ising or Rotor models \cite{sachdev99}. As an example, the ferromagnetic-paramagnetic PT in the presence of a transverse magnetic field that occurs in the crystals of LiHOF\textsubscript{4} and CoNb\textsubscript{2}O\textsubscript{6} can be modelled using Ising models. Similarly, an O(3) quantum rotor model describes the transition from paramagnetic to an ordered antiferromagnetic phase in the TiCuCl\textsubscript{3} crystal while increasing the pressure. Phase transitions in disordered systems like spin glasses can also be studied using ising/rotor models\cite{sachdev99}.

The Landau-Ginzburg-Wilson \cite{hohenberg15} method is a commonly used approach in QPT analytical studies. However, one disadvantage of analytical techniques is that they must use perturbative methods to analyze field theories, which might fail in tightly coupled or substantially disordered systems. Such phase transitions could be studied using a variety of other approaches, including static and dynamic large-N calculations, conformal field theory, perturbative renormalization group, the local moment technique, and numerous numerical methods. In particular, the numerical renormalization group (NRG) approach \cite{moca10,lee05} has made substantial development. QPTs can also be investigated utilizing mean-field methods such as DMFT \cite{kotliar06,vollhardt12}.
\subsubsection{Quantum Simulation of Quantum Phase Transition}

Quantum Phase Transition is a complex phenomenon to study, and in many cases, it is not well known. Quantum simulation offers a relatively new way to understand the phenomenon. However, there is little work in this area, and most of it is based on analog simulations. Ultra-Cold atom simulations was able to capture the Superfluid to Mott-insulating transition due to their inherent advantage in simulating the Hubbard model \cite{greiner2002}. Further, Dynamical Quantum Phase Transitions (DQPT) observed in topological insulators are also simulated using Ultra-cold atoms\cite{flaschner16}. In addition, a protocol for accessing Loschmidt amplitudes in systems of Ultra-Cold atoms has just been proposed \cite{daley12,pichler13}. A dynamical phase transition was also observed in ion-trap simulations of the Transverse field Ising model (TFIM). A protocol for capturing the Metal-Insulator phase transition occurring in topological insulators with ultra-thin films is presented in \cite{ju14}. Also, in \cite{peng2005} an NMR-quantum computer was used to simulate the ground state phase transition in the Heisenberg chain. 

This area has limited resources for Digital quantum simulations. However, DQS was able to capture the phase transition occurring in the Transverse Field Ising Model \cite{powers21}. Through TFIM, we can model quantum phase transitions in a variety of crystal types.

\subsection{Light-matter interaction}
For more than a century, light-matter interactions have been a developing field within physics~\cite{loudon2000quantum}. Though since exotic coupling regimes have been introduced, light-matter interactions have had a rebirth in recent years. A different path has arisen that gives fresh views to the field, notably light-matter interactions in the ultra-strong and deep-strong coupling regimes~\cite{forn2019ultrastrong,frisk2019ultrastrong}, in addition to the exciting theoretical and practical advances in this area in which quantum technologies already have a number of applications~\cite{nielsen2002quantum}. The light-matter coupling in the ultra-strong coupling regime is of the order of the electromagnetic mode frequency and smaller and in the case of deep-strong coupling regime, is larger than the mode frequency. Variational approach for simulating ultrastrong light-matter coupling is explored in~\cite{di2020variational}. 

In this section, we provide an overview of quantum simulations of light-matter interactions, focusing on three paradigmatic examples: the Jaynes Cumming model, the quantum Rabi model describing the coupling between two-level quantum system and a quantized bosonic mode, and the Dicke model.

\subsubsection{Closed Quantum systems}
Closed quantum systems are isolated quantum systems, which are defined as systems that do not exchange information (energy and/or matter) with other systems. In contrast to an open system, in which an atom in free space is coupled to multiple modes of the EM field and an initial excitation in the system is irreversibly emitted from the atom into the field modes, never to be reabsorbed by the atom; in a closed system, such as the Jaynes–Cummings model, the excitation is periodically transferred between the atom and the field~\cite{ficek2014quantum}. The state of the system is represented by $\psi(t) \in \mathcal{H}$, where $\mathcal{H}$ denotes the Hilbert space of the system. The time evolution of the system is governed by the Schrödinger's equation,
\begin{equation}
	i\hbar\frac{\partial \psi{(\textbf{r},t)}}{\partial t} = \hat{H} \psi{(\textbf{r},t)}
\end{equation}

where $\hat{H}$ is the Hamiltonian of the system. This formulation can alternatively be expressed in terms of a density matrix. The time evolution of the density matrix is given by,
\begin{equation}
	\rho(t) = U \rho(0) U^{\dagger}
\end{equation}
where $\rho(t)$ is the density matrix of the system and $U$ is the time evolution operator. This operator for a time-independent Hamiltonian can simply be written as,
\begin{equation}
	U = \exp(-i\frac{\hat{H}t}{\hbar})
\end{equation}
This conforms to the well-known quantum mechanical approaches in which a system's evolution is unitary. Because all operations are unitary, such systems can be directly simulated on a quantum computer, as discussed in greater detail in Section\ref{2} and Section \ref{4}. We know, however, that no system is actually isolated enough for this to hold true. As a result, quantum systems are open rather than closed. We'll discuss about open quantum systems and how to simulate them in the section \ref{5.3}. We now discuss on some examples for quantum simulation of light matter interaction.

\subsubsection{Jaynes Cummings Model}
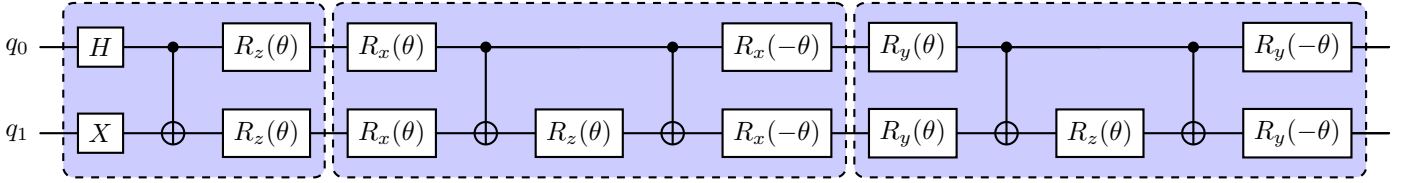
\begin{figure}
\begin{center}
	\begin{tikzpicture}
		\node[scale=1] {
			\begin{quantikz}
				\lstick{$q_0$} & \gate{H}\gategroup[2,steps=3,style={dashed, rounded corners,fill=blue!20, inner xsep=2pt}, background]{{\sc }} & \ctrl{1} & \gate{R_z(\theta)} & \gate{R_x(\theta)}\gategroup[2,steps=5,style={dashed, rounded corners,fill=blue!20, inner xsep=2pt}, background]{{\sc }} & \ctrl{1} & \qw & \ctrl{1} & \gate{R_x(-\theta)} & \gate{R_y(\theta)}\gategroup[2,steps=5,style={dashed, rounded corners,fill=blue!20, inner xsep=2pt}, background]{{\sc }} & \ctrl{1} & \qw & \ctrl{1} & \gate{R_y(-\theta)} & \qw
				\\
				\lstick{$q_1$} & \gate{X} & \targ{} & \gate{R_z(\theta)} & \gate{R_x(\theta)} & \targ{} & \gate{R_z(\theta)} & \targ{} & \gate{R_x(-\theta)} & \gate{R_y(\theta)} & \targ{} & \gate{R_z(\theta)} & \targ{} & \gate{R_y(-\theta)} & \qw
			\end{quantikz}
		};
	\end{tikzpicture}
	\caption{Quantum circuit for implementing Jaynes Cumming Model~\cite{Bind2020JC}. The initial state is represented by the first portion, which is followed by two $Rz$ gates that represent the atomic and number operators and $\theta$ depends on the time step. The interaction terms $\hat{\sigma_+}\hat{b}$ and $\hat{\sigma_-}\hat{b^\dagger}$ are described in the last two parts of the circuit.
	}
	\label{fig:JC_model}
\end{center}
\end{figure}

The interaction of a two-level atom with a single-mode quantized field under the rotating-wave approximation, is known as Jaynes-Cummings (J-C) model. It is one of the simplest models in quantum optics. It is widely used in cavity and circuit quantum electrodynamics, as well as quantum information processing. A quantum harmonic oscillator can represent the quantized EM field~\cite{bose2014all}, and Rabi frequency characterises the coupling between atom and field~\cite{fink2008climbing}. The Hamiltonian of J-C model~\cite{cummings1965stimulated} is given by 
\begin{equation}
	\hat{H}_{JC} = \hbar \omega \hat{b}^\dagger \hat{b} + \frac{\omega_0}{2} \hat{\sigma_z} + \hbar\lambda(\hat{\sigma_+}\hat{b} + \hat{\sigma_-}\hat{b}^\dagger)
\end{equation}
where $\hat{b}$ and $\hat{b}^{\dagger}$ are the bosonic annihilation and creation operators respectively and $\omega$ is the frequency of the quantized bosonic field. In second term, $\hat{\sigma_z}$ is the atomic operator of two-level system and $\hat{\sigma}_+$ \& $\hat{\sigma}_-$ represent the atomic raising and lowering operator respectively. In~\cite{Bind2020JC}, the simulation of the J-C model utilising a quantum simulator and actual quantum hardware was demonstrated. The generated Rabi frequency graphs are obtained and found to be almost similar to the theoretical values. The Holstein-Primakoff (H-P) transformation~\cite{holstein1940field} is used to map the Bosonic operators to the Spin operators (for mapping methodologies, see section \ref{3}), with the simulation being executed in three steps: \textit{first}, initial state preparation, \textit{second} unitary decomposition, and \textit{third}, measurement. Based on ~\cite{Bind2020JC}, fig.~\ref{fig:JC_model} shows a quantum circuit model of the complete system.
\end{multicols}

\begin{multicols}{2}
\subsubsection{Quantum Rabi Model}
The quantum Rabi model~\cite{rabi1936process} is the fundamental paradigm of light-matter interactions. It is made up of a two-level spin that is coupled to an electromagnetic bosonic mode. The Dicke model~\cite{dicke1954coherence} is the consequence of extending it to numerous spins linked to the same mode. In the Schrödinger picture $(\hbar = 1)$, the Hamiltonian for the quantum rabi model is $\hat{H}_{QRM}$.
\begin{equation}
	\hat{H}_{QRM} = \frac{\omega_0^D}{2} \hat{\sigma_z} + \omega^D \hat{b}^\dagger \hat{b} + g \hat{\sigma_x}^(\hat{b}^\dagger +\hat{b})
\end{equation}
where $\omega_0^D$ is the spin frequency, $\omega^D$ is the bosonic mode frequency, $g$ is the light matter coupling, $\hat{b}$, $\hat{b}^\dagger$ are annihilation and creation operators of the electromagnetic bosonic mode, and $\hat{\sigma}_{x,z}$ are Pauli matrices. The quantum simulation of the quantum rabi model in various coupling regimes is investigated in multiple quantum architectures. In~\cite{braumuller2017analog}, a method for analog quantum simulation of the quantum Rabi model in ultrastrong and deep strong coupling regimes employing an orthogonal two-tone drive to the qubit is suggested, using a circuit QED system. This protocol's experimental analysis may be found in~\cite{leppakangas2018quantum}. 

In a circuit QED setting,~\cite{mezzacapo2014digital} presents a digital-analog quantum simulation of the Rabi model in a superconducting circuit. They demonstrated how to encode the Rabi model Hamiltonian using a Jaynes-Cummings interaction in a superconducting setting. The quantum Rabi Hamiltonian may be split into two parts with a Jaynes Cumming (rotating) and an anti-Jaynes Cumming (counter-rotating) term. In conventional circuit QED setups with faster control of qubit frequencies, these two terms can be executed digitally. Authors in~\cite{pedernales2015quantum, lv2018quantum} suggest a quantum simulation of the Rabi model in trapped ion systems.
\subsubsection{Dicke Model}
The Dicke model~\cite{dicke1954coherence}, which is a natural extension of the quantum Rabi model~\cite{rabi1936process}, is made up of $N$ two-dimensional spins that are connected to a single electromagnetic field bosonic mode. The Hamiltonian for the Dicke model~\cite{mezzacapo2014digital} is given by $(\hbar = 1)$, 
\begin{equation}
	\hat{H}_{D} = \sum_{i=1}^N \frac{\omega_0^D}{2} \hat{\sigma_z}^i + \omega^D \hat{b}^\dagger \hat{b}+ g \sum_{i=1}^N \hat{\sigma_x}^i(\hat{b}^\dagger +\hat{b})
\end{equation}
An analog quantum simulation of the Dicke model is presented by~\cite{aedo2018analog}, in trapped ion systems. It is feasible to induce a coupling between qubit and phonon states by irradiating ions with laser beams, allowing the models of interest to be replicated. By carefully selecting rabi frequency and detuning \& phase for red and blue-sideband, the form of the Dicke model may be approximated using the sum of the red and blue-sideband Hamiltonians of the ion system. 

The digital-analog quantum simulation of the Dicke model in the superconducting circuit is shown in~\cite{mezzacapo2014digital,lamata2017digital, remizov2018analog}. The Dicke Hamiltonian can be simulated by breaking it down into digital steps via unitary decomposition~\cite{mezzacapo2014digital}. The protocol in~\cite{mezzacapo2014digital}  can be implemented by combining collective single-qubit rotations with collective Tavis-Cummings dynamics. Using the results from~\cite{berry2007efficient,wiebe2011simulating}, the authors demonstrated that the quantum resources required to simulate the Dicke Hamiltonian with an error less than $\epsilon$ scale effectively with the number of spins $N$ and excitation permitted in the bosonic mode $M$. Given adequate coherence and low gate errors, the Dicke model may thus be explored. Since collective single-qubit addressing is possible, this quantum simulation technique is suitable for circuit QED. In ~\cite{lamata2017digital}, an exploration of a digital-analog quantum simulator of generalized Dicke models in superconducting circuits along with the a framework for the quantum simulation of the biased and pulsed Dicke models, and numerical simulations for all light-matter interaction regimes, with a detailed error analysis is presented.
\end{multicols}

\begin{figure}
\centering
\begin{quantikz}
	\lstick{1} & \ctrl{3}\gategroup[4,steps=15,style={dashed, rounded corners,fill=blue!20, inner xsep=2pt, inner ysep=8pt}, background]{{\sc }} & \gate{Z} & \ctrl{2} & \gate{Z} & \ctrl{1} & \gate{Z} & \gate{X} & \gate{R} & \gate{X} & \gate{Z} & \ctrl{1} & \gate{Z} & \ctrl{2} & \gate{Z} & \ctrl{3} & \qw \\
	\lstick{2} & \qw & \qw & \qw & \qw & \targ{} & \qw & \qw & \qw & \qw & \qw & \targ{} & \qw & \qw & \qw & \qw & \qw \\
	\lstick{3} & \qw & \qw & \targ{} & \qw & \qw & \qw & \qw & \qw & \qw & \qw & \qw & \qw & \targ{} & \qw & \qw & \qw \\
	\lstick{4} & \targ{} & \qw & \qw & \qw & \qw & \qw & \qw & \qw & \qw & \qw & \qw & \qw & \qw & \qw & \targ{} & \qw
\end{quantikz}
\caption{Quantum circuit for implementing the first local beam-splitting interaction term~\cite{quantum2010013}, where $R = \exp(-i \frac{\epsilon_{+-}}{8}\hat{\sigma_z}^{(1)})$, and parameters $\epsilon_{+-}$ represents the two modes.}
\label{fig:beam_split}
\end{figure}
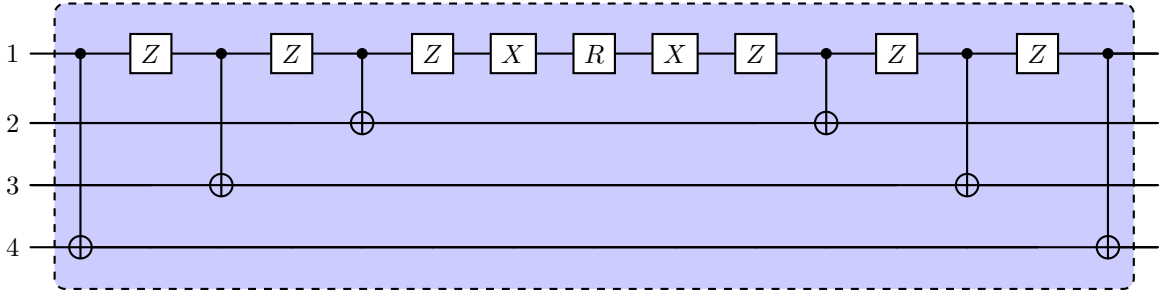

\begin{multicols}{2}
	
\subsubsection{Optical elements}
The digital quantum simulation of linear and nonlinear optical elements in superconducting circuits were reported in~\cite{quantum2010013}. A digital quantum simulation approach was suggested that encodes generic multi-mode bosonic interactions into a sequence of single-qubit and two-qubit gates via a boson-qubit mapping. There are three steps to the method: mapping bosons to qubits, Suzuki-Trotter decomposition, and gate decomposition. A few examples based on this approach was discussed, including boson sampling and boson sampling Hamiltonian, beam-splitters, sequences of beam splitters, two-mode squeezing, and Bogoliubov transformations. A mesh of $M(M-1)/2$ beam splitters and phase shifters governs the unitary evolution of the Boson Sampling Hamiltonian of the output photon-number statistics of a linear-optics network with $M$ photonic modes~\cite{quantum2010013}. The first beam-splitter of the mesh would only require $2$ modes $\times$ $2$ qubits per mode i.e. 4 qubits, because the initial state of boson sampling contains only one photon per mode. As a result, the beam splitter unitary may be expressed as a product of $8$ local unitaries. The circuit breakdown of the first local unitary reported in~\cite{quantum2010013} in detail, is presented in fig. \ref{fig:beam_split} above.

In~\cite{encinar2021digital}, a demonstration on digital quantum simulation of single-mode and two-mode squeezing and beam-splitter interaction Hamiltonians in IBM quantum devices was shown. It was based on the idea of combining Trotter~\cite{lloyd1996universal} and gate-decomposition techniques~\cite{somma2002simulating} with boson-qubit mappings~\cite{somma2005quantum} to encode generic bosonic hamiltonians into a sequence of single-qubit and two-qubit gates. The single mode squeezing was considered to be unitary and represented it as a product of local unitaries and then decomposed local unitaries in terms quantum gates. Same method is used for beam splitter and two mode squeezing. The achieved fidelities exceeded 90\% for modest single-mode squeezing and fidelities ranging from 60\% to 90\% for two-mode interactions such as two-mode squeezing and beam-splitting, depending on the parameters. In the case of two-mode interactions, quantum circuits are substantially more sophisticated and incorporate many more two-qubit gates. Digital quantum simulation of continuous-variable quantum information processing~\cite{gao2018programmable,zhang2019engineering} and some phenomena commonly examined in analog quantum simulators, such as the Casimir effect~\cite{johansson2013nonclassical, wilson2011observation} and molecular force fields~\cite{olivares2017quantum, cao2019quantum}, are some of the examples of applications of quantum simulation of optical components.

\subsection{Open Quantum systems}\label{5.3}
In general, quantum systems are never completely isolated from their environments, and they must be considered as open systems for accurate study. Open quantum systems OQS have always been of great interest to scientists and researchers, as more realistic systems that we usually encounter are open systems. It's crucial to understand the dynamics of open quantum systems if one wants to explore a variety of quantum phenomena~\cite{breuer2002theory, gardiner1991quantum, stolze2008quantum}. Although, there has been considerable work in the literature on the simulation of the dynamics of open quantum systems, practical implementation has only lately received interest due to their complexity and requirements for large memory and time resources~\cite{barreiro2011open}. The theory of open quantum systems~\cite{breuer2002theory,rivas2012open} provides a useful framework for understanding a wide range of phenomena, including quantum decoherence~\cite{vacchini2016quantum}, out of equilibrium many-body dynamics~\cite{PhysRevA.94.052120}, quantum field theory~\cite{CLIFTON20011}, and information processing in physical systems.

The open systems can be considered as a composite systems having two parts-, a subsystem S (usually referred as the system) and a bath B (environment). The Hilbert space of composite system can be describe by the tensor product of the individual Hilbert spaces of two subsystems~\cite{lidar2019lecture}. 
\begin{equation}
	\mathcal{H} = \mathcal{H}_S \otimes \mathcal{H}_B 
\end{equation}
where $\mathcal{H}_S$ represents the Hilbert space for the system and $\mathcal{H}_B$, the Hilbert space for the environment. Usually, the dimensions of the Hilbert space for the environment (B) $d_B \to \infty$ while for the system (S), is finite dimensional~\cite{lidar2019lecture}. 

\subsubsection{Dynamics of open quantum systems}
We assume that the total system evolves according to the Schrödinger equation and that it is described by the density matrix $\rho(t)$, which in general, cannot be written as a product of the density matrices of the subsystems~\cite{lidar2019lecture}. Such subsystems are called correlated. We are more interested in system S, so partial trace is used to effectively average out the effect of components of B from the combined density matrix~\cite{lidar2019lecture,gupta2020optimal}. For any wave function $|\psi\rangle$,  the solution of Schrödinger equation is given as

\begin{equation}
	|\psi(t)\rangle = U|\psi(0)\rangle = \exp(-i\frac{H}{\hbar}t)|\psi(0)\rangle        
\end{equation}
In terms of density matrix formalism,
\begin{equation}
	\rho(t) = U\rho(0)U^\dag
\end{equation}
While dealing with the open quantum systems, a common assumption made is that the density matrix of total system can be written as~\cite{gupta2020optimal}, 
\begin{equation}
	\rho(t) = \rho_S \otimes \rho_B.        
\end{equation}
The state of the system then evolves as,
\begin{equation}
	\rho_S = tr_B\{U(t) \rho_S \otimes \rho_B U^\dag(t)\}.
\end{equation}
Such a evolution can be representation in a variety of ways. One of them, the operator sum represented is described below. Because the conditions of trace preservation and complete positivity can be clearly defined, the operator sum representation provides a convenient method to describe these types of mapping~\cite{bacon2001universal}. This maps any initial state $\rho_S(0)$ to its final state using a dynamical map $\Phi_t: S(H_S) \to S(H_S)$, such that 
\begin{equation}
	\rho_S(0) \to \rho_S(t) = \Phi_t\rho_S(0).
\end{equation}
$\Phi_t$ is completely positive and can be represented in terms of Kraus operators as, 
\begin{equation}
	\Phi A = \sum_i \Omega_i A \Omega_i^\dag,
\end{equation}
for operator A, where $\Omega_i$ are Kraus operators and $\sum_i \Omega_i^\dag\Omega_i = I_S$. 
There are two types of open quantum dynamics: Markovian and Non-Markovian. In~\cite{bacon2001universal} and~\cite{sweke2016digital}, a study on possible methods for simulating the Markovian and Non-Markovian quantum dynamics respectively, of open quantum systems is presented. 
\subsubsection{Quantum Simulation of Open Quantum Systems}

Although it is commonly established that arbitrary unitary operations can be used to simulate the dynamics of closed quantum systems, the same cannot be said for more generalized operations that correspond to the dynamics of open quantum systems~\cite{bacon2001universal}. In the literature, there has been some work on quantum simulation of open systems. Digital quantum simulation of open quantum systems has gained attention only recently. The IBM Q-Experience simulator was used to simulate open quantum dynamics incorporating amplitude damping and depolarisation, as well as Pauli channels and collision models, in~\cite{garcia2020ibm}. Quantum channels have also been studied using discrete time quantum algorithms, in~\cite{hu2020quantum}. In this section we review some algorithms for the digital quantum simulation of the dynamics of open quantum systems.

Theoretical understanding of the dynamics of open quantum systems has been a significant focus of research in the past decade. Only recently the actual implementation of quantum algorithms for these simulations were possible, thanks to cloud services such as IBM Q-Experience. An approach for simulating the most general quantum dynamics, and dynamical semi-groups of quantum channels, which describe Markovian dynamics: continuous-time processes resulting from interactions with a Markovian environment in the Born approximation, of an open quantum system on a quantum computer is presented in~\cite{bacon2001universal}. The authors theoretically demonstrated how the simulation of Markovian dynamics can be simplified to the construction of generators for a Markovian semigroup and how the generators of Markovian semigroups transform under the composition procedures of linear combination and unitary conjugation. For a large system, however, obtaining the Markovian semigroup generator can be problematic. Authors in~\cite{wang2011quantum} present an algorithm for modelling the Markovian dynamics of an open quantum system. The Hamiltonian of the open system, the operators by which the open system interacts with the environment, the spectral density of the environment, and temperature are all inputs to the method. As a result, one does not explicitly address the master equation characterising dynamics. The environment is replicated in the simulation by a set of ancillary qubits that are meant to have the same effect on the open system as the simulated environment. The energy spectrum of the system could be obtained by monitoring the response of these ancillary qubits as they interact with the open system. Another quantum algorithm is presented in~\cite{sweke2014simulation} for the simulation of arbitrary Markovian dynamics of a qubit, described by a semigroup of single-qubit quantum channels ${T_t}$ specified by a generator $\mathcal{L}$, with a goal to find a quantum circuit, acting on only the system qubit and a single-ancilla qubit and using at most poly single-qubit and CNOT gates, that approximates the superoperator $T_t = \exp(t\mathcal{L})$ up to the chosen accuracy. They generalise into the superoperator regime recombination results based on higher order Suzuki-Lie-Trotter formulae~\cite{suzuki1990fractal} and apply ~\cite{wang2011quantum}'s work for the building of circuits for the modelling of the semigroups required by~\cite{bacon2001universal}.

The Stinespring dilation theorem~\cite{stinespring1955positive} directly suggests a simple methodology for open quantum systems simulation, in which one introduces an initially pure state environment with a size square of the system size (in general), in order to simulate the open system dynamics of the system via Hamiltonian dynamics of the larger system-environment combination. In~\cite{gupta2020optimal} the Stinespring dilation theorem~\cite{stinespring1955positive} is used to create quantum circuits for simulation of Markovian and Non-Markovian evolution of the dynamics of OQS. The circuits replicate the environment using ancilla qubits, and memory effects are produced by storing system information on extra qubits. They proposed a three step method for the digital quantum simulation of Markovian and Non-Markovian dynamics of OQS: $i)$ writing the Kraus representation of the dynamics, ii) using Stinespring dilation theorem to find a unitary operator for the evolution of system and environment, and iii) decomposing the unitary operator into elementary gates to create the quantum circuit for the dynamics. Simulating Non-Markovian systems is based on the idea of retaining some knowledge about the system in the ancillas. This can be accomplished by adding additional information-storing environment qubits. The environment is partially traced out at the end of each step, and SWAP gates are used to update the information stored on qubits for the following step. This makes it possible to simulate memory effects in dynamics. They numerically simulate the amplitude damping channel and dephasing channel as examples of the framework and infer (Non-) Markovianity from the dynamics' (non-) monotonic behaviour.

In~\cite{hu2020quantum}, a general quantum algorithm for evolving open quantum dynamics on quantum computing devices based on the Sz.-Nagy theorem~\cite{langer1972b} was shown. The Sz.-Nagy dilation, which is a version of the Stinespring dilation theorem, requires a significantly smaller dimension increase and can save significant processing resources. It guarantees that the Kraus operators driving the time evolution of the density matrix can be transformed into unitary matrices with minimal dilation. Furthermore, the starting state can be evolved using unitary quantum gates while requiring far fewer resources than traditional Stinespring dilation as in~\cite{gupta2020optimal}. The algorithm's implementation on IBM Q platform was shown.

The unitary decomposition of operators is another way for simulating the dynamics of open quantum systems. On the basis of it,~\cite{schlimgen2021quantum} presents a general algorithm for implementing the action of any non-unitary operator on any arbitrary state on a quantum device. It is shown that any quantum operator may be decomposed exactly as a linear combination of at most four unitary operators and demonstrated this approach on a two-level system with zero and finite temperature amplitude damping channels, on IBM Q's qasm simulator. Based on Quantum Imaginary Time Evolution (QITE)~\cite{kamakari2022digital} and Modified Stochastic Schrodinger Equation (MSSE)~\cite{jo2022simulating}, two more ways of implementing non-unitary operators on quantum computing devices have recently been presented. In~\cite{kamakari2022digital} two techniques based on QITE for digital quantum simulation of the dynamics of open quantum system was presented. The first approach entails converting the Lindblad equation to a Schröndinger type equation having unitary (implemented using trotterization) and non-unitary (implemented using QITE) terms, while the second entails creating an ansatz for the density operator. In the second algorithm too, QITE is used to implement the non-unitary part. The MSSE is used to demonstrate a general approach for simulating many-body open quantum systems on quantum computing devices~\cite{jo2022simulating}. Unitary Hamiltonian evolution (implemented naturally using trotterisation) and non-unitary Lindblad evolution are the two parts of MSSE approach. It is demonstrated that there is an optimal quantum circuit arising from the non-commutativity of Hamiltonian and Lindblad operators, which allows the simulation to run at a relatively large discretised time interval, which is helpful for quantum simulations with constrained resources. The method was demonstrated on IBM Q devices for a two-level system in a heat bath and a dissipative transverse field Ising model. Experimental and numerical results of dissipative transverse field Ising model are presented in~\cite{jo2022simulating, kamakari2022digital}. The implementation of the decay operator can be achieved by the circuit shown below. For more details, refer section \emph{'METHOD'} from~\cite{jo2022simulating}.

\begin{center}
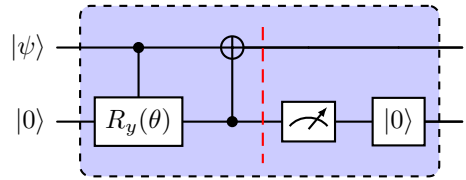

	\begin{quantikz}
		\lstick{$|\psi\rangle$} & \ctrl{1}\gategroup[2,steps=4,style={dashed, rounded corners,fill=blue!20, inner xsep=2pt, inner ysep=8pt}, background]{{\sc }} & \targ{}\slice{} & \qw & \qw & \qw \\
		
		\lstick{$|0\rangle$} & \gate{R_y(\theta)} & \ctrl{-1} & \meter{} & \gate{\ket{0}} & \qw
	\end{quantikz}
	\captionof{figure}{Quantum circuit for implementing the decay operator for dissipative transverse field Ising model~\cite{jo2022simulating}.}
\end{center}

In ~\cite{jo2022simulating}, implementation of a quantum contact process (QCP) was presented. 
Recently, an algorithm (Quantum (Q)-TEDOPA) based on the Time Evolving Density operator with Orthogonal Polynomials Algorithm (TEDOPA) for simulating the non-perturbative dynamics of open quantum systems was proposed and demonstrated in~\cite{guimaraes2022efficient}. This method is based on a precise transformation of the environment and Hamiltonians of system-environment interaction. The simulation of an open quantum system is mapped to a closed Hamiltonian dynamics problem that can be efficiently simulated on a quantum computer.

The digital quantum simulation of open quantum systems became feasible after quantum computing devices became accessible through cloud services such as IBM-Q Experience, IonQ platform, and the algorithms suggested in the literature may finally be put to the test. The versatility of contemporary quantum computing resources was demonstrated in~\cite{garcia2020ibm} by implementing a range of open quantum system models on IBM Q devices. The authors use a parameterized family of completely positive and trace preserving (CPTP) maps to characterise the dynamics of OQS. Different types of open quantum system dynamics can be simulated by modifying the parameter $(0 \leq p \leq 1)$. This work shows that the near term quantum computing devices can be a versatile test platform for the theory of open quantum systems. The evolution of hard probes in the quark-gluon plasma (QGP) can be modelled as an open system evolving in a hot medium. In~\cite{de2021quantum}, a framework based on the Stinespring dilation theorem~\cite{stinespring1955positive} was proposed, for the evolution of hard probes in the QGP as a Lindblad equation and study how simulations using noisy intermediate scale quantum (NISQ) devices might help progress theoretical investigations of hard probes in the QGP. In~\cite{gupta2020digital}, a theoretical framework for digital quantum simulations of ENAQT (environment-assisted quantum transport) in photosynthetic Fenna–Mathews–Olson systems (FMO) was proposed. They demonstrate a digital simulation of the FMO complex as an example.
\end{multicols}

\begin{table}[ht]
\centering
\begin{tabular}{ |p{2.2cm}|p{5cm}|p{5cm}|p{2cm}|  }
	\hline
	\textbf{Algorithm} & \textbf{Method} & \textbf{Example studied} & \textbf{Reference}\\ [1ex]
	\hline
	Algorithm 1 & Stienspring Dilation theorem & - & ~\cite{gupta2020optimal} \\ [1ex]
	\hline
	Algorithm 2 & Sz.-Nagy theorem & - & ~\cite{hu2020quantum} \\ [1ex]
	\hline
	Algorithm 3 & Quantum Imaginary Time Evolution (two approaches presented) & Two level system in a heat bath \& Dissipative TFIM & ~\cite{kamakari2022digital} \\ [6.5ex]
	\hline
	Algorithm 4 & Unitary decomposition of operators & Two level system & ~\cite{schlimgen2021quantum} \\ [3ex]
	\hline
	Algorithm 5 & Modified Stochastic Schrödinger Equation & Quantum contact process \&  Dissipative TFIM & ~\cite{jo2022simulating} \\ [4ex]
	\hline
	Algorithm 6 (Q-TEDOPA) & Time Evolving Density operator with Orthogonal Polynomials Algorithm (TEDOPA) &  Exciton transport between two electromagnetically coupled molecules of the photosynthetic Fenna-Matthews-Olsen (FMO) complex in green sulphur bacteria & ~\cite{guimaraes2022efficient} \\ [4ex]
	\hline
\end{tabular}
\caption{Various algorithms for simulating the dynamics of open quantum systems.}
\end{table}

\begin{multicols}{2}
Some variational approaches to find the steady state of an open quantum systems are also proposed in literature. In~\cite{liu2021variational} a variational quantum algorithm for determining the steady state of open quantum systems was proposed. The authors use parameterized quantum circuits (PQC) in this technique to prepare the purification of the steady state, which may be evaluated efficiently with quantum circuits. They employed the swap test to estimate the cost function, which was defined using the Frobenius norm of Lindbladian. Then they optimise the quantum circuit's parameters to find the steady state. The method was demonstrated by numerically simulating the dissipative TFIM using QuTip~\cite{johansson2012qutip}. A variational framework based on Variational Quantum Monte-Carlo methods and neural network representation of the density matrix for the simulation of non-equilibrium steady states in open quantum systems is proposed in~\cite{nagy2019variational}. The density matrix was parameterized using a neural network approach, and they used a stochastic reconfiguration strategy for parameter optimization, which was proven to approximate the system's real-time dynamics. The Metropolis-Hasting technique can be used to find the expected value of any observable once the optimal parameters have been found. They demonstrated this Variational Monte Carlo approach to investigate the steady-state properties of the dissipative spin-1/2 XYZ model.

Beyond applications in open quantum system dynamics, the approaches outlined in the previous discussion have general applicability to problems in quantum chemistry and physics in the realm of quantum computing. The decomposition provided in~\cite{schlimgen2021quantum} can be used to implement Hamiltonians or dipole operators, which are prevalent in quantum chemistry and are generally non-unitary. Where perturbative approaches fail to provide solutions, the algorithm suggested in~\cite{guimaraes2022efficient} can be extended to Hamiltonian dynamics simulations of generic quantum biological systems~\cite{christensson2012origin,mohseni2014quantum} and condensed matter systems~\cite{wang2021polariton,vasilevskiy2004electron}. The results of~\cite{gupta2020digital} can be used in ENAQT's digital quantum simulation.

\subsection{Quantum Chemistry}
In quantum chemistry, simulation of the problems can save a lot of time and money invested in experiments. Theoretical chemistry allows us to solve the equations governed by the fundamental laws of physics including those of quantum mechanics. These analytical solutions can be used to engineer products that have vast applications worldwide. However, there might be cases in which it is impossible to solve the equations using known methods. In that case, computational chemistry comes into the picture which solves the problem with the help of numerical methods. There have been a lot of techniques developed to numerically simulate the dynamics of the molecules. Widely used methods are the Density Functional Method which uses DFT, wavefunction approach, Embedding methods and Diagrammatic method \cite{ref_2motta2021emerging}.  

Quantum chemistry involves the study of the dynamics of different molecular systems at the quantum level. The size of the molecules is perfect enough to manifest quantum mechanical phenomena. These systems can be solved by using the Schrödinger equations if the potential of the system is known. Simulating quantum systems are a difficult task for classical computers. With the improvement of the quality of the description of the Schrödinger equation, the computational resources grow exponentially. A lot of research is ongoing for developing algorithms which require less computational power and give better accuracy. Despite that, as the molecular size increases, the available computational requirement explodes \cite{ref_3sherrill2010frontiers}.

Quantum computation promises many advantages in simulating quantum systems \cite{ref_4webber2022impact}. This is the idea of simulating a quantum system on another quantum system without calculating classical ones. Quantum computers can provide a faster and more accurate simulation of various molecular systems and allow a researcher to calculate properties like reaction rates \cite{ref_5lu2011simulation}, molecular vibrations \cite{ref_6mcardle2019digital}, ground state and low-lying excited state energies \cite{ref_7wang2008quantum} etc.

Initially, the Quantum Phase Estimation algorithm (explained in section \ref{QPE}) was thought to have the potential for solving many quantum chemistry problems. It can be used to solve quantum chemistry problems related to the time evolution of the molecular Hamiltonian and estimate the Eigen-energy value at a particular time. It must be noted that the computational resources required by the QPE algorithm to have accurate results require noise resilient quantum computers and a relatively large number of qubits. Hence, it has been considered to be a long-term quantum algorithm \cite{ref_9elfving2020will}. Currently, in this NISQ era, algorithms like Variational Quantum Eigen-solver and hybrid quantum machine learning methods have higher promise as they do not need a large number of qubits and can be applied on NISQ devices \cite{ref_11andersson2022quantum}.

In the following section, we discuss some of the quantum chemistry problems and parameters that can be calculated with the help of quantum simulation and hence benefit from the development of quantum computing technologies. These include Ground state and Excited state simulation, Molecular designing simulation, Spectral analysis and Chemical Reaction simulation.

\subsubsection{Ground and Excited state simulation}

Calculation of the electronic structure is one of the most important problems in computational chemistry. It is the problem which studies the properties of the many electron systems interacting with external potential and the Coulomb potential. If the simulation of the electronic structure of a molecule can be done efficiently then it can fasten the progress of development of the industries like pharmaceutical, chemical engineering and many other areas of chemistry. Calculation of energy levels is a part of the electronic structure problem. This section focuses on the ground state and excited state simulation of the molecules and recent development in quantum computing algorithms for the same.

There are many ways to simulate the energy states of a molecule. But before that, it is important to describe the Hamiltonian of the system and map it to the Hamiltonian of qubits which is in terms of products of Pauli matrices. We can take the example of the hydrogen atom. To get the Hamiltonian, we need to first count the number of orbitals in the molecule. For hydrogen there are 2 1s orbitals and oxygen has 1s, 2s, 2$p_{x}$, 2$p_{y}$ and 2$p_{z}$. This totals up to 14 orbital if one includes spin state as well. Now, in the case of H\textsubscript{2}O, it can be considered that the molecule has 2 unoccupied orbitals with the largest energy, leaving us with a total of 12 occupied orbitals states with spin. Then the second quantisation method is used to formulate the Hamiltonian \cite{ge_1lanyon2010towards} (for brief discussion of second quantisation see section \ref{Quantum simulation Hubbard Mode}). The equation of Hamiltonian becomes as shown in eq. \ref{second-q}

\begin{equation}\label{second-q}
	\hat{H}=\sum_{i, j=1}^{12} h_{i j} \hat{a}_{i}^{\dagger} \hat{a}_{j}+\frac{1}{2} \sum_{i, j, k, l=1}^{12} h_{i j k l} \hat{a}_{i}^{\dagger} \hat{a}_{j}^{\dagger} \hat{a}_{k} \hat{a}_{l}
\end{equation}

In the above equation, $\hat{a}^\dagger$ and $\hat{a}$ are the Fermionic creation and annihilation operators. The one-body and two-body interaction terms are given by  $h_{ij}$ and $h_{ijkl}$. These are in the form of integrals (shown in eq. \ref{h_ij} and eq. \ref{h_ijkl}) and can be calculated using numerical integration methods. 

\begin{equation}\label{h_ij}
	h_{i j}=\int d \vec{r}_{1} \chi_{i}^{*}\left(\vec{r}_{1}\right)\left(-\frac{1}{2} \nabla_{1}^{2}-\sum_{\sigma} \frac{Z_{\sigma}}{\left|\vec{r}_{1}-\vec{R}_{\sigma}\right|}\right) \chi_{j}\left(\vec{r}_{1}\right)
\end{equation}

\begin{equation}\label{h_ijkl}
	h_{i j k l}=\int d \vec{r}_{1} d \vec{r}_{2} \chi_{i}^{*}\left(\vec{r}_{1}\right) \chi_{j}^{*}\left(\vec{r}_{2}\right) \frac{1}{r_{12}} \chi_{k}\left(\vec{r}_{2}\right) \chi_{l}\left(\vec{r}_{1}\right)
\end{equation}

This fermionic Hamiltonian can be mapped to Hamiltonian as the product of the Pauli matrix using Bravyi-Kitaev transformation already discussed in \ref{BK Transformation}.  One can also use parity basis and particle and spin conservation method to further reduce the number of qubits needed. 

In quantum computation, there are simulation methods which can be used to simulate and calculate the energy states of the molecule using its Hamiltonian. The phase estimation method using Trotterisation in which the Eigen-energy values are encoded into the phase of the propagator is one method. Also, another two ways for simulations are Direct implementation of Hamiltonian using first-order \cite{ge_3daskin2018direct} and second order Trotterisation \cite{ge_4daskin2018generalized}. Yet another method is the Direct-measurement method. These algorithms are all Phase Estimation Algorithm (PEA) type algorithms. The most useful method in the NISQ era of quantum computers is the Variational Quantum Eigen Solver \cite{ge_6mcclean2016theory}. The paper \cite{ge_7bian2019quantum} shows that the VQE method requires the least number of qubits for scaling. While PEA type of methods is shown to have higher accuracy just by one measurement but they require a large number of qubits. This shows that the VQE algorithm is best suited for NISQ era quantum computers while PEA types are the ones best suited for long term quantum computers.\\

\vspace{1em}
\hrule
\vspace{2em}
\noindent
\textbf{Number Operator}
\begin{center}
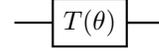

	\begin{tikzpicture}
		\node[scale=1]{
			\begin{quantikz}
				\lstick{} & \gate{T(\theta)} & \qw
			\end{quantikz}
		};
	\end{tikzpicture}
	\captionof{figure}{Here $T(\theta)$ is the phase gate such that $T(\theta)\ket{0}=\ket{0}$ and $T(\theta)\ket{0}=\exp(-\iota\theta)\ket{1}$}
	\label{fig:my_label}
\end{center}
\vspace{2em}
\noindent
\textbf{Number-excitation operator}\\
\begin{center}
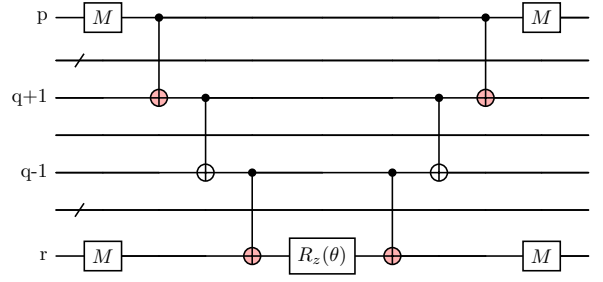

	\begin{tikzpicture}
		\node[scale=0.75]{
			\begin{quantikz}
				\lstick{p} & \gate{M} & \ctrl{2} & \qw & \qw & \qw & \qw & \qw & \ctrl{2} & \gate{M}  & \qw \\
				& \qw\qwbundle{} & \qw & \qw & \qw & \qw & \qw & \qw & \qw & \qw & \qw \\
				\lstick{q+1} & \qw & \targ[style={fill=red!30}]{} & \ctrl{2} & \qw & \qw & \qw & \ctrl{2} & \targ[style={fill=red!30}]{} & \qw & \qw \\
				& \qw & \qw & \qw & \qw & \qw & \qw & \qw & \qw & \qw & \qw \\
				\lstick{q-1} & \qw & \qw & \targ{} & \ctrl{2} & \qw & \ctrl{2} & \targ{} & \qw & \qw & \qw \\
				& \qw\qwbundle{} & \qw & \qw & \qw & \qw & \qw & \qw & \qw & \qw & \qw \\
				\lstick{r} & \gate{M} & \qw & \qw & \targ[style={fill=red!30}]{} & \gate{R_z(\theta)} & \targ[style={fill=red!30}]{} & \qw & \qw & \gate{M} & \qw
			\end{quantikz}
		};
	\end{tikzpicture}
	\captionof{figure}{M gate is the combined set of $\{H,Y\}$ gates taken in order ($Y=R_x(-\frac{\pi}{2})$)}
	\label{fig:my_label}
\end{center}
\vspace{2em}
\noindent
\textbf{Double excitation operator}\\
\begin{center}
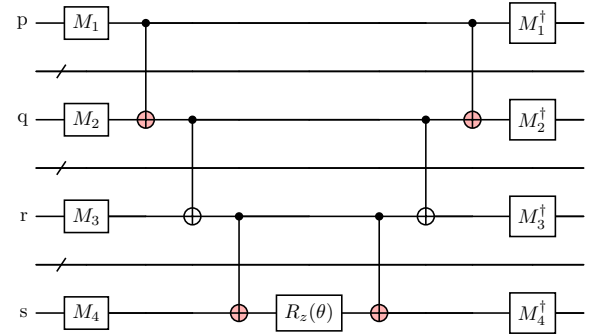

	\begin{tikzpicture}
		\node[scale=0.75]{
			\begin{quantikz}
				\lstick{p} & \gate{M_1} & \ctrl{2} & \qw & \qw & \qw & \qw & \qw & \ctrl{2} & \gate{M_1^\dagger}  & \qw \\
				& \qw\qwbundle{} & \qw & \qw & \qw & \qw & \qw & \qw & \qw & \qw & \qw \\
				\lstick{q} & \gate{M_2} & \targ[style={fill=red!30}]{} & \ctrl{2} & \qw & \qw & \qw & \ctrl{2} & \targ[style={fill=red!30}]{} & \gate{M_2^\dagger} & \qw \\
				& \qw\qwbundle{} & \qw & \qw & \qw & \qw & \qw & \qw & \qw & \qw & \qw \\
				\lstick{r} & \gate{M_3} & \qw & \targ{} & \ctrl{2} & \qw & \ctrl{2} & \targ{} & \qw & \gate{M_3^\dagger} & \qw \\
				& \qw\qwbundle{} & \qw & \qw & \qw & \qw & \qw & \qw & \qw & \qw & \qw \\
				\lstick{s} & \gate{M_4} & \qw & \qw & \targ[style={fill=red!30}]{} & \gate{R_z(\theta)} & \targ[style={fill=red!30}]{} & \qw & \qw & \gate{M_4^\dagger} & \qw
			\end{quantikz}
		};
	\end{tikzpicture}
	\captionof{figure}{\centering In the circuit,
		$(M_1,M_2,M_3,M_4)=\{(H,H,H,H),(Y,Y,Y,Y),\newline(H,Y,H,Y),(Y,H,Y,H),(Y,Y,H,H),(H,H,Y,Y),\newline(Y,H,Y,H),(H,Y,H,Y)\}$}.
	\label{fig:my_label}
\end{center}
\vspace{1.5em}
\noindent
\textbf{Excitation Operator}
\begin{center}
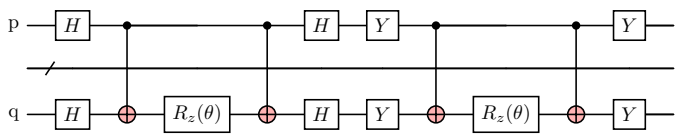

	\begin{tikzpicture}
		\node[scale=0.75]{
			\begin{quantikz}
				\lstick{p} & \gate{H} & \ctrl{2} & \qw & \ctrl{2} & \gate{H} & \gate{Y} & \ctrl{2} & \qw & \ctrl{2} & \gate{Y} & \qw \\
				& \qw\qwbundle{} & \qw & \qw & \qw & \qw & \qw & \qw & \qw & \qw & \qw & \qw \\
				\lstick{q} & \gate{H} & \targ[style={fill=red!30}]{} & \gate{R_z(\theta)} & \targ[style={fill=red!30}]{} & \gate{H} & \gate{Y} & \targ[style={fill=red!30}]{} & \gate{R_z(\theta)} & \targ[style={fill=red!30}]{} & \gate{Y} & \qw
			\end{quantikz}
		};
	\end{tikzpicture}
	\captionof{figure}{Y gate is nothing but $Y=R_x(-\frac{\pi}{2})$}
	\label{fig:my_label}
\end{center}
\vspace{1.5em}
\noindent
\textbf{Coulomb and exchange operators}\\
\begin{center}
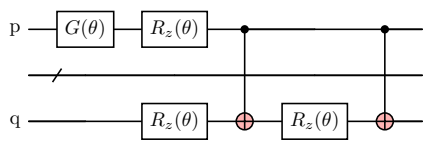

	\begin{tikzpicture}
		\node[scale=0.75]{
			\begin{quantikz}
				\lstick{p} & \gate{G(\theta)} &\gate{R_z(\theta)} & \ctrl{2} & \qw & \ctrl{2}  & \qw \\
				& \qw\qwbundle{} & \qw & \qw & \qw & \qw & \qw \\
				\lstick{q} & \qw & \gate{R_z(\theta)} & \targ[style={fill=red!30}]{} & \gate{R_z(\theta)} & \targ[style={fill=red!30}]{} & \qw
			\end{quantikz}
		};
	\end{tikzpicture}
	\captionof{figure}{$G(\phi)$ is the global phase gate which is expressed as $\exp{(-i\phi)}1$}
	\label{fig:my_label}
\end{center}
\begin{center}
	\vspace{1.5em}
	\hrule
	\noindent
	\emph{\textbf{Notation:}}\\
	\vspace{0.5em}
	\begin{tikzpicture}
		\node[scale=0.75]{
			\begin{quantikz}
				& \ctrl{3} & \qw & \ctrl{3} & \qw \\
				& \qw & \qw & \qw & \qw \\
				& \qw & \qw & \qw & \qw \\
				& \targ[style={fill=red!30}]{} & \gate{} & \targ[style={fill=red!30}]{} & \qw 
			\end{quantikz}
			= \begin{quantikz}
				& \qw & \ctrl{1} & \qw & \qw & \qw & \qw & \qw & \ctrl{1} & \qw \\
				& \qw & \targ{} & \ctrl{1} & \qw & \qw & \qw & \ctrl{1} & \targ{} & \qw \\
				& \qw & \qw & \targ{} & \ctrl{1} & \qw & \ctrl{1} & \targ{} & \qw & \qw \\
				& \qw & \qw & \qw & \targ{} & \gate{} & \targ{} & \qw & \qw & \qw
			\end{quantikz}
		};
	\end{tikzpicture}
\end{center}
\hrule

\subsubsection{Molecular designing simulation}
Being able to study the dynamics of the molecules and their time evolution allows scientists to design and come up with new molecules which can be used as products in the market or as a treatment for certain diseases. The last 2 years of the Covid-19 pandemic indicate the importance of speeding up these processes of designing molecules. These can be solved using two methods. The first approach is using the Born-Oppenheimer approximation. Alternatively, the dynamics of the quantum molecular systems can be expressed as the simple product of time-dependent electronic and nuclear wave functions \cite{ref_mol_dyn7abedi2010exact}. Simulation using these methods requires a higher computational cost. Although, the computational cost can be lowered by making certain approximations it often increases the errors \cite{ref_mol_dyn8curchod2018ab,ref_mol_dyn9gatti2014molecular}.

Quantum dynamics has its relevance in the study of non-equilibrium processes with potential energy surfaces, dynamics of molecular and solid state systems with electron and nuclear dynamics and optimal quantum control theory. Quantum optimal control theory is of high interest. It is nothing but the theory of controlling the dynamics of quantum systems using external lasers. Applications are immense and growing rapidly \cite{ref_mol_dyn3koch2019quantum}. The theory has been experimentally verified with bond dissociation experiments \cite{ref_mol_dyn4damrauer2002control}, isomerisations \cite{ref_mol_dyn5vogt2005optimal} and molecular fragmentation \cite{ref_mol_dyn6daniel2003gonza}.  

This field has shown vast growth in recent years. Although, current QC algorithms are often used for demonstration purposes only for which only simple molecules are considered. This is because the current state-of-the-art quantum computers are limited in terms of qubits. These confines the algorithms to BO approximations and do not allow the inclusion of non-adiabatic effects \cite{ref_mol_dyn10ollitrault2021molecular}. 

In this field of work, the most famous class of quantum algorithms that are used is Variational Quantum Algorithms (VQAs). These algorithms use a hybrid approach, that is, the simulation of the system is done on a parameterised quantum circuit whose parameters are optimised classically using some cost function. The authors of \cite{ref_mol_dyn11gao2021computational} use Variational Quantum Eigensolver to simulate their molecular system. They specifically study the rearrangement dynamics of the molecule. One reason why VQE is often used is that the current era of quantum computers is noisy, but VQAs are adaptable to the noisy nature of QCs if one uses Hardware efficient Ansatz \cite{ref_mol_dyn12motta2021emerging}. In \cite{ref_mol_13gao2021quantum} a hybrid method for calculating and designing Deuterated High-Efficiency Organic Light Emitting Diodes was proposed. They use machine learning methods to calculate the Ising model systems and then followed by the implementation of the VQE algorithm to calculate the quantum efficiency of the molecular system to obtain the optimal Deuterated molecule.  
Apart from using VQAs, one can also use the Digital Quantum Simulation method for molecular dynamics. It has been used for laser-driven systems. The work \cite{ref_mol_14magann2021digital} describes the use of the theory of quantum optimal control to simulate the dynamics of molecules. Although their approach is also hybrid they do not use a variational approach. The steps involve mapping to the qubits, and Hamiltonian simulation followed by the qubit readout. To find the optimal control field, the readout states of the qubit are used by the classical computer for optimisation. The optimisation function can be decided based on the Quantum optimal control theory. This approach can be used for controlled bond stretching, the orientation of dipole-dipole coupled (OCS) rotors and preparing the state of the light-harvesting complex in a controlled manner.
\subsubsection{Spectral analysis}

Spectral analyses refer to the study of spectral properties. These properties include the spectrum of frequencies of vibrations and related quantities like eigenvalues and energies. It is well known that matter can never be at rest. At the quantum level, even the tightly bonded molecules execute oscillations. More commonly these oscillations are called vibrations. One can study vibrations in time independent and time dependent picture. The former allows us to perform spectral calculations like Infrared and Raman spectroscopy \cite{spect_1zhu2009theoretical} and fluorescence \cite{spect_2wang2014franck} which have importance in determining solar cells performance \cite{spect_4debbichi2012vibrational} and industrial dyes \cite{spect_5dhananasekaran2016adsorption}. The dynamics of vibrations have much more applications including the dynamics of reactions \cite{spect_8antoniou2001internal} and electronic transport \cite{spect_10hwang2004harmonic}. Also, these affect large frequency temporal resolved laser experiments \cite{spect_6mcardle2019digital}.

This is a field of great importance. There have been methods to accurately simulate the systems but they are limited to a few particle systems. One such method is Real-space, grid-based method. When the simulation of the molecular systems is done on a classical computer, we are restricted to using a finite basis for spaning the infinite dimensional Hilbert space. The full configuration interaction or FCI method can provide accurate solutions for electronic structures but scales up exponentially with the increase in system size \cite{spect_11helgaker2014molecular}. This field uses configurational methods discussed in \ref{section_conf}.

\end{multicols}
\begin{center}
\begin{longtable}[!htp]{ |p{3cm}|p{3cm}|p{3cm}|}\hline
	\textbf{Operator Name} &\textbf{Symbol} &\textbf{Circuit} \\\hline
	Number Operator & $h_{pp}a^{\dagger}_pa_p$ & circuit 1
	\\\hline
	Excitation Operator & $h_{pq}(a^{\dagger}_pa_q+a^{\dagger}_qa_p)$ & circuit 2 \\\hline
	Coulomb and exchange operators & $h_{pqqp}a^{\dagger}_pa^{\dagger}_q+a_qa_p)$ & circuit 3\\\hline
	
	Number-excitation operator & $h_{pqqr}(a^{\dagger}_pa^{\dagger}_q+a_qa_r+a^{\dagger}_ra^{\dagger}_q+a_qa_p)$ & circuit 4\\\hline
	
	Operator for double excitation & $h_{pqrs}(a^{\dagger}_pa^{\dagger}_qa_ra_s+a^{\dagger}_sa^{\dagger}_ra_qa_p)$ & circuit 5\\\hline
	
	\caption{Quantum circuits for second quantisation operators. The circuits are presented above}
\end{longtable}
\end{center}

\begin{multicols}{2}

\subsubsection{Chemical Reaction simulation}

The analytical results in quantum chemistry are very important when it comes to understanding a chemical reaction. They allow us to know the steps and mechanisms involved in a chemical reaction \cite{ref_rxn1engkvist2018computational}. Again, classical computers pose the problem of fewer computational resources. Solving the Schrodinger equation and simulating its time evolution requires an exponential increase in the size of the system. Also, increasing the degree of freedom requires us to have more size in the system. 

Quantum Computers as already known, can easily simulate or can be used to propagate the Schrödinger equation. They show a promise of completing the same task in polynomial time \cite{ref_rxn2aaronson2005quantum} when compared to classical computers. Although the limited number of qubits and noisy hardware makes things difficult for simulating accurate results. Still, there are noise mitigation techniques which can be used to reduce the noise. Most of the algorithms in Quantum computation for chemical reactions are based on the approach of Digital quantum simulations \cite{ref_rxn3salathe2015digital}. Specifically, there have been examples of reactions which are controlled and driven by the external laser fields. 
\end{multicols}
\begin{figure}[h]
\centering
\includegraphics[width=0.5\textwidth]{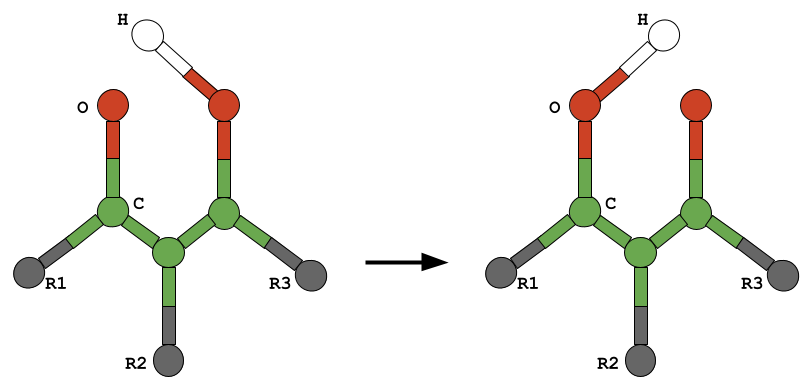}
\caption{Figure shows the isomerisation of substituted malonaldehyde which is non-symmetric. Authors of \cite{ref_rxn4lu2011simulation} performed the digital quantum simulation of taking this molecule into consideration. A double well potential has been considered during isomerisation.}
\label{Img-3}
\end{figure}
\begin{multicols}{2}

The DQS based approach can be found in \cite{ref_rxn5halder2018digital}. Authors of both have studied the isomerisation reactions in a double welled potential and showed the time evolution of the reactant and product states. The former implemented the algorithm on an NMR based quantum computer, while the latter simulated the same in the quantum simulator of IBM, called ibmq\_qasm\_simulator. The theoretical approach remains the same in both the cases which are to find the Hamiltonian in second quantisation and then to the qubit system. The former uses a GRAPE technique \cite{ref_rxn7rabitz2000whither} to implement the pulses on the NMR QC to implement the unitary operations. This technique provides an efficient implementation on the NMR quantum computers.

\subsection{Bioinformatics}

There has been a lot of development of algorithms and mathematical tools to solve biological problems. Bioinformatics is one of the filed of utmost importance as it provides solutions to a better lifestyle, helps fight against widespread diseases and much more. This field explores complex areas like human genome, modelling biomolecule’s behaviour in different environments, calculating binding affinity of a ligand etc. These problems can widely be categorised into three subsections. These are Drug Discovery, Genome Sequencing and Proteomics. 

The research in the field of finding better computational methods has been tremendous. Some of these methods are used in alignment of sequence \cite{bioinfo_2apostolico1998sequence,bioinfo_3zuker1991suboptimal}, computational genetics \cite{bioinfo_4marcotte2000computational}, data processing for X-ray diffraction \cite{bioninfo_6prescher2015dioptas}. These problems can be solved using computational methodologies. But the current computational resources are not enough to simulate large bio-molecules.  Therefore, there has been a shift in attention from classical computers to quantum computers as a computational platform. The following subsections explain three major problems that can be found in the field of Bio-informatics and also describes the emerging role of Quantum computation.

\subsubsection{Drug Discovery}

The Discovery of drugs were used to happen accidentally like penicillin. But now with advancements in technology drug discovery has become not a random process but a process involving steps and procedures. Various chemical compounds are selected from the database and extensively searched if they can be a potential drug. Also, their synthesisability is studied. Following this, the compound is optimized to maximise the affinity and then it is gone through trials including animals succeeded by human trials.

The development of drugs is a very long process and consists of the following stages including target discovery then molecular design followed by pre-clinical studies and lastly clinical trial. This makes the process of creating a marketable drug expensive and time-consuming.  Pharmaceutical research has been using high-throughput screening technology Discovering drugs involves searching through target-binding compounds. This is a long process and even expensive. While the computational methods for molecular docking can help to identify the molecules which bind with the target. 

The computational accuracy of the docking search depends on the description of the compound in the software’s library used for the process. This software is often called the docking engine. The docking methods used by these engines can vary from software to software. Some common examples of these methods are Auto-Dock Vina \cite{drug_1cosconati2010virtual}, MedusaDock \cite{drug_2wang2019medusadock} and Glide \cite{drug_3halgren2004glide}. These approaches mostly try to properties of the compounds and receptors which can bind together. The accurate results are given by the density functional theory. As usual, the classical computational methods are limited to small sized molecules and receptors \cite{drug_4li2021drug}.

The quantum algorithm for Quantum machine learning is a faster and cheaper solution to the problem of classical computations. The most promising area of quantum algorithms for this field is the Quantum Generative Adversarial Network \cite{drug_5li2021quantum}. Quantum GAN with Hybrid Generator \cite{drug_6li2021quantum} is one of the QGAN algorithms. It consists of a parameterised quantum circuit for finding the feature vector and followed by a classical DNN to predict the bond matrix in terms of graphs for the drug. In this method, there are various variations. One is Patched-QGAN-HG \cite{drug_7huang2021experimental}. Apart from QGAN based approaches, there are Image search based methods. These methods involve the quantum convolution neural network \cite{drug_8henderson2020quanvolutional}. These are based on convolution neural networks. In QNN the filters are replaced with the quantum circuits. The quantum variational autoencoders have been tried to perform the drug simulations but have not yet shown any better performance than classical VAE. VAE is a method which is based on probabilistic search \cite{drug_4li2021drug}.

\subsubsection{Genome sequencing}
Genome sequencing is the process of figuring out the order of nucleotides of the DNA. A DNA consists of an order of genetic letters As, Cs, Gs and Ts. The human genome is made up of over 3 billion letters. To understand a genome, its sequencing is very important. This will allow scientists to find genes much more quickly as they contain information on where the genes are. The study of genome sequence has great significance to scientists to understand how they direct the growth and maintenance of a full organism.

Traditional computational methods use De Novo assembly \cite{genome_1laserson2011genovo} to construct an original DNA from an unstructured set of reads without using any pre-requisite knowledge like DNA sequence length, composition etc.  The complexity depends on the size of the genome. As an example, it takes nearly 50 hours for the human genome on a supercomputer. This time might be acceptable for research tasks but is not fit for the case of emergencies. These assembly tools are based on Overlap Layout Consensus (OLC) algorithm \cite{genome_2myers2005fragment}. It uses an OLC graph in which the vertices are presented by a read while the overlap between any two reads corresponds to the vertices of the graph. Then the Hamiltonian path is found which is the path that consists of all the edges and each vertex is visited only once giving the original genome. 

Quantum computers specifically quantum annealing can solve this problem. Since it is a graph problem, it can be formulated into a QUBO formulation. The OLC-graphs are converted to the QUBO/Ising model \cite{genome_3glover2018tutorial} which is then embedded in the quantum annealing system and then as an output one is given the Hamiltonian path. Apart from QUBO formulation, the QAOA method is used for DNA sequencing to accelerate the de-novo sequencing. 

\subsubsection{Proteomics}
Proteomic is a field that has started merging with quantum technologies. This field studies the electronic structure of the proteins in a given cellular system of any organism. Proteomics is defined as a group of proteins present in the organism. This allows scientists to study the properties of proteins like energy levels, dipole moments, amino acid charges, their electric potential and a plethora of many things. This field emerged after the human genome project (HGP) was completed in 2001 \cite{p_1hubbard2002functional}. HGP involved the identification of more than 30,000 genes in humans, which gave way to the study of the proteins expressed by the genes. The problems involved in this field include the characterisation of protein structures, inter protein interaction (interactome) and phosphorylation (phospho-proteomics). 
Protein folding is one problem which comes up after the identification of proteins. One needs to study the proteins to disclose the knowledge of how the proteins are encoded in genes. Classical algorithms of protein folding can sample small conformation space. Many quantum algorithms have been proposed to solve this problem. The paper \cite{p_2robert2021resource}, proposes a hamiltonian and the variational quantum algorithm for folding a polymer chain with N monomers on a lattice, specifically for 10 amino acid Angiotensin worked out on 22 qubits. Gate based algorithms have also been proposed as in paper \cite{p_3khatami2022gate}. This field has a great potential for growth with quantum computers offering an exploration of large conformational space for protein folding. 
\section{Error Mitigation}\label{6}
Near term quantum computers are not fault-tolerant. The two most significant hurdles to scalable universal quantum computers are error sensitivity and noise. Errors can arise in each quantum computation step, making it difficult for efficient digital quantum simulations. The errors can be broadly classified as (i) State preparation errors, (ii) Gate errors and (iii) Measurement or Readout errors. The gate errors are further classified into Coherent and Incoherent errors. The coherent errors preserve the purity of the state. It typically occurs due to miscalibration in the control parameters. Now, one can understand incoherent errors in two ways. Either it can be considered coherent errors with randomly varying control parameters or an operation that entangle the qubit with the environment. State Preparation and Measurement errors are sometimes together addressed as SPAM errors. Compared to gate errors, SPAM errors occur only at the beginning or end of the circuit and do not accumulate with increasing circuit depth.

There are two proposals for achieving Fault-tolerant quantum computation. The first method uses non-abelian quasiparticles called anyons in the topological matter to perform error-free quantum computation (Topological Quantum Computation) \cite{kitaev2003}. Another approach for fault-tolerant quantum computation is using Quantum Error-Correcting (QEC) codes  \cite{simon13}. One can use these codes to detect and remove gate errors during computation. While the first proposal is still in its infancy, the second approach requires computational resources unattainable in near-term devices. For instance, using Surface code, a ubiquitous QEC code, one needs millions of physical qubits to perform the fault-tolerant computation of Shor’s algorithm \cite{endo18}. 

Readout errors make approximately $15\%$ of error in quantum computation (Superconductor qubit based). Thus mitigating readout errors holds importance. A straightforward approach is using the operator rescaling method for error mitigation. It uses the documented readout errors to correct the results via post-processing. But it cannot mitigate the correlated errors in the computation. Another approach to minimize the readout errors is the calibration matrix method. 

\emph{Calibration Matrix Method}: In this method, before each time evolution experiment, we perform a calibration experiment to characterize the device for each of the $2^{N}$ basis states. We organize the results of each calibration experiment in a $2^{N} \times 2^{N}$ matrix after the calibration experiment. Each member of the matrix $P ij$ represents the probability of a system preparing in-state $i$ and measuring in-state $j$. By applying the inverse of this matrix to the noisy results, we would get results with mitigated measurement errors. While applying this method, we need to make two crucial assumptions; 1) The readout error is caused by classical noise, and 2) the noise is independent of the quantum gates applied to the system. A recent work \cite{maciejewski20} shows that readout errors in quantum systems based on superconducting qubits can be effectively explained using simply classical noise models. Further, to prevent the exponential scaling of the calibration matrix with system size, we assume the error due to noise is local and correlates to a subset of qubits \cite{kandala17,sun20}. Then the error model is called tensored error model, and the calibration matrix will be the tensor product of $2\times 2$ noisy matrice. Also, sometimes due to strong noise, the inverse of the calibration matrix will not be well defined. In such a scenario, we have to find the Moore-Penrose pseudo inverse of the calibration matrix. Tensored error models do not address the errors due to cross-talk between qubits during readout. Therefore, recently a measurement error mitigation scheme that addresses cross-talk errors was proposed \cite{bravyi20}.

Next, let's move on to the mitigation of gate errors. As mentioned earlier, there are Coherent and Incoherent gate errors. Incoherent errors are usually modelled as depolarizing noise. There exist methods to mitigate depolarizing noise \cite{urbanek21,vovrosh21}. Coherent errors are more damaging than incoherent ones. But in \cite{wallman16} it was shown that one could convert coherent errors to incoherent errors through randomized compiling. Thus in principle, the coherent errors also can be mitigated. But there are other approaches to mitigating gate errors, including the popular one called Zero Noise Extrapolation (ZNE) \cite{temme17, Li17}. We will discuss some of the schemes to reduce the gate errors in the following section.

\emph{Zero Noise Extrapolation}: A ZNE consists of two steps Noise Scaling and Extrapolation. In noise scaling, we would intentionally scale up the noise level of the quantum circuit. Noise scaling can be done in two ways. The first approach is called time scaling or pulse stretching. In this method, we stretch the control pulses and thereby increase the noise in the circuit. The second approach is called the unitary folding. Here we map the unitary operation $U$ to $U(U^{\dagger}U)^{n}$, where n is an integer. Thus in the quantum circuit, it would increase the circuit depth and thereby scale the noise. The unitary folding can be applied globally (Circuit Folding) or locally (Gate Folding) \cite{giurgica20}. Using Noise scaling, we would calculate the expectation value of the observable (that we want to measure) at different noise levels. Once we have the regression between the expectation value (of the observable) and noise, we can evaluate the expectation value at the zero-noise limit through extrapolation. The model used for performing extrapolation depends upon the noise model assumed. Polynomial extrapolation is generally used in the weak noise limit if the number of data points ($m$) is equal to or greater than $d+1$, where $d$ is the degree of the polynomial. Of the two variants of polynomial extrapolation, linear extrapolation is used when $d=1$ and Richardson extrapolation when $d=m-1$ \cite{temme17}. The polynomial extrapolation is inefficient when there is a low error rate and a large number of noisy gates. In such cases, we need to resort to other extrapolation methods such as poly-exponential extrapolation \cite{giurgica20} and exponential extrapolation \cite{endo18}. 

\emph{Probabilistic Error Correction}:   The Probabilistic Error Cancellation (PEC) \cite{temme17} works based on two ideas. The first idea is the quasi-probability representation of the ideal gates. It essentially means that we should represent an ideal gate as a linear superposition of noisy quantum gates \cite{pashayan15}. The real coefficients in such a representation form a quasi-probability distribution, i.e., the sum of the coefficients will be normalized but differ from the standard probabilities by taking negative values. Using quasi-probability representation, any observable represented using an ideal gate set can be translated into a noisy gate set representation. One could directly find the expectation values of noisy gates from the hardware. Using it, we could derive the ideal expectation value of any observable. Unfortunately, this strategy demands the execution of a large number of circuits, which rises exponentially with circuit depth and is often impractical. Thus we use the second idea of probabilistically sampling from the quasi-probability representation. The Monte Carlo average that follows would give the approximate expectation value of the observables. 

\emph{Other methods for error mitigation}: Methods like PEC and ZNE discussed above require complete knowledge of the underlying noise model to be efficient. In most cases, experimentalists only have imperfect knowledge of the noise model. Therefore people are working on learning-based QEM techniques using ZNE and PEC to repress errors via an automatic learning process. Examples of such methods include \emph{Clifford Data Regression} \cite{czarnik21}, and \emph{Learning-based quasi-probability} method \cite{strikis20}. In addition, there is also another approach based on Gate Set Tomography for error mitigation without being noise aware \cite{endo18}. Apart from the popular ones, there are other methods for error mitigation. Examples of such methods include \emph{Dynamic Decoupling} \cite{zhang14}, \emph{Quantum Subspace Expansion} \cite{mcclean17,colless18}, \emph{Stochastic error mitigation} \cite{sun21} and so on. Most of the methods we discussed do not use ancilla qubits. Another class of error mitigation methods also uses ancilla qubits for error mitigation. One ubiquitous example is the \emph{Stabilizer-based (Symmetry verification) error mitigation}. It uses ancilla to perform measurements on conserved quantities of the problem Hamiltonian, such as spin or parity. Any error would change conserved quantities indicated upon ancilla measurement (similar to stabilizers in QEC). This method is often used in variational state preparation methods \cite{mcardle19b,sagastizabal19b}. Another error mitigation method that utilizes ancilla qubit is the recently proposed \emph{Virtual Distillation} \cite{huggins21,czarnik21b}. It suppresses noise by simultaneously preparing numerous copies of a noisy state, entangling and measuring them to virtually prepare a more pure state than the original noisy one. Virtual distillation is quite promising with its exponential suppression of error rate.

Another class of errors that we haven't discussed yet is the Algorithmic errors. Compared to others, these errors are not of physical origin. One ubiquitous example is the trotterization errors arising in the Hamiltonian simulation. The prevalent method for mitigating trotterization error is the ZNE. We perform the noise scaling using small trotter steps and then apply extrapolation. Recently, another approach that exploits symmetry of the system to mitigate trotterization error was proposed \cite{tran21}. It is a symmetry protection technique that causes destructive interference of the errors arising in each trotter step. An extensive introduction to algorithmic and other error mitigation techniques is provided in \cite{endothesis}.

\section{Software tool sets}\label{7}

Quantum computing is a field which merges many disciplines. At its current age, it is at a stage where it has evolved to enter the industry. Many start-ups in quantum computing have come up in recent years including Xanadu, IonQ, Zapata, PASQAL, just to name a few. These start-ups are now collaborating with bigger organisations providing them quantum solutions to the existing problems. Over the years cloud solutions have become very famous in this field. Many companies have been building software which allows one to execute their problems with real quantum hardware without the need to learn quantum computing. This section provides the details of many software \& online platforms for quantum simulation encountered during the survey. Their details have been given in the table 2.
\end{multicols}
\begin{center}
\begin{longtable}[!htp]{ |p{2.5cm}|p{2.5cm}|p{4.5cm}|p{4.5cm}|  }\hline
	\textbf{Software Package} &\textbf{Domain} &\textbf{Tasks} &\textbf{Noteworthy attributes} \\\hline
	QuASeR \cite{soft_1sarkar2021quaser}  &Bioinformatics &DNA Sequencing &One can perform DNA sequencing using the de-novo assembly on gate based and quantum annealers. Uses TSP, QUBO, Hamiltonian methods and QAOA algorithms \\\hline
	InQuanto \cite{soft_2greene2022modelling,soft_4kirsopp2021quantum}&Chemistry &Ground and Excited states,Spectroscopy, Molecular Geometry,Transition Pathways, Reaction Pathways, Ligand Protein Binding, Molecular Forces &Mostly uses VQE methods and its variations like ADAPT-VQE, Penalty VQE, VQ Deflation, Quantum Subspace Expansion and Iterative Qubit-excitation Based VQE for computation of the tasks. The packages also comes with Error Mitigation techniques like PMSV and SPAM \\\hline
	MQS \cite{soft_4kirsopp2021quantum}  &Chemistry &Computation of Solubility, Viscosity, Partition coefficient values, Phase equilibria calculations for vapour-liquid-, liquid-liquid and solid-liquid-equilibria. &Maps the models of quantum chemistry models (DFT, PMx,COSMO-RS/SAC, GNFx-xTB) to Quantum computer hardware through cloud based methods. Allows submitting calculations which are accessed and pipelined to further steps. Has applications for process design, product design and Material Design \\\hline
	OpenFemion \cite{mccleanOF} &Chemistry \& Condensed Matter &Computation of Trotter error operators, symbolic Fourier transformation, preparing fermionic Gaussian states, routines for generating Hamiltonians of the Hubbard model, the homogeneous electron gas (jellium), general plane wave discretizations, and d-wave models of superconductivity and wide range of data structures important in Quantum chemistry &Everything from efficient data structures for encoding fermionic operators to fermionic circuit primitives for use on quantum devices is included in the package. \\\hline
	Fermionic.jl &Chemistry \& Condensed Matter &Fermionic operators can be constructed both in the full Fock space or in the fixed particle number subspace, can be used to perform fermionic quantum computation. Compute average particle number, one body matrices entropy, partially traced systems , majorization relations, m-bodies density matrices, one body entropies and more. &Julia tool kit for fermionic simulations and fermionic quantum computation \\\hline
	MISTIQS \cite{powers21} &Condensed Matter &Translation of circuits into executable circuit objects for IBM, Rigetti, and Google quantum devices, domain-specific IBM and Rigetti compilers developed for TFIM simulations, support for user-defined time dependence functions for external fields, full XYZ model support in Hamiltonian constructions. &A full-stack, cross-platform software for creating, constructing, and running quantum circuits for simulating quantum many-body dynamics of Heisenberg Hamiltonians-governed systems. \\\hline
	QuSpin \cite{weinberg17,weinberg19} &Condensed Matter &Can Implement Exact diagonalisation, Lanczos Algorithm, Floquet Hamiltonian simulation of a wide range of many-body system. Also have parallel computing capabilities &An open-source Python package that supports the use of various (user-defined) symmetries in one and higher dimensions, as well as (imaginary) time evolution following a user-specified driving protocol, for exact diagonalization and quantum dynamics of arbitrary boson, fermion, and spin many-body systems. \\\hline
	Kwant \cite{groth14} &Condensed Matter &Calculation of transport parameters (conductance, noise, scattering matrix), dispersion relations, modes, wave functions, different Green's functions, and out-of-equilibrium local values, other computations involving tight-binding Hamiltonians &Kwant is a Python package for computing numerical quantum transport. It  provide a user-friendly, ubiquitous, and high-performance toolkit for simulating physical systems of any dimensionality and geometry that can be characterised by a tight-binding model. \\\hline
	ArQTIC \cite{bassman21b} &Condensed Matter &Dynamic Simulation, QITE Simulation, can simulate materials that can be modeled by any Hamiltonian derived from a generic, one-dimensional, time-dependent Heisenberg Hamiltonain &An open-source, full-stack software package built for the simulations of materials on quantum computers \\\hline
	
	Quantavo~\cite{feito2008quantavo} & Quantum Optics & A framework which can declare, manipulate and characterize quantum states of light (finite number of modes, and finite dimensional), and implement linear optics circuits such as Beam Splitters (BS), Phase Shifters (PS), arbitrary unitary transformations of the modes etc. & A Maple toolbox for linear optics and quantum information in Fock space \\\hline
	QuantumOptics.jl ~\cite{kramer2018quantumoptics} &Open quantum systems & numerical simulation of the dynamics of OQS, finding the steady-state of OQS \& time correlation functions, Optimizes processor usage and memory consumption &A Julia framework for simulating open quantum systems \\\hline
	HOQST: Hamiltonian Open Quantum System Toolkit ~\cite{Chen2022} &Open quantum systems & simulating the dynamics of OQS, supports various master equations, as well as stochastic Hamiltonians & A Julia toolkit for simulating the open quantum system dynamics \\\hline
	Mitiq \cite{larose21} &Error Mitigation &Zero-noise extrapolation, Probabilistic error cancellation, and Clifford data regression &Python package for error mitigation on noisy quantum computers \\\hline
	QuaEC \cite{criger13} &Error Correction &Support for maniuplating Pauli and Clifford operators, as well as binary symplectic representations and automated analysis of error-correcting protocols based on stabilizer codes &Python library for working with quantum error correction and fault-tolerance \\\hline
	CHP: CNOT-Hadamard-Phase \cite{aaronson2004} &Error Correction &Construct quantum error-correction designs and debug them. Numerically study massive, highly entangled quantum systems. Generate random stabilizer quantum circuit, Shor 9-qubit quantum error-correcting code &High-performance simulator of stabilizer circuits (Quantum Error Correction) \\\hline
	QuaSiMo \cite{nguyen21} &Hybrid quantum-classical algorithms &Dynamicalsimulation, VQE, Symmetry reduction, Fermion qubit mapping, QITE, QAOA & A composable design scheme for the development of hybrid quantum/classical algorithms and workflows for applications of quantum simulation \\\hline
	QuEST \cite{jones19b} &-&Many functions for simulating decoherence, Calculating density inner product, Hilbert Schmidt distance, Purity, Fidelity, many quantities from Density matrix &simulator of quantum circuits, state-vectors and density matrices. \\\hline
	qsims &-&qsims represents the spatial wavefunction of a particle as a discretized wavefunction on a grid, Internal states of the particle can be represented using multiple grid, Potentials and couplings between the internal states can be specified, and potentials can be position- and state-dependent. &A tool for studying quantum computing using optical lattices, General-purpose quantum simulation software package, capable of simulating the dynamics of systems with a wide range of Hamiltonians, qsims is not limited to optical lattices, and could be adapted for use in many other physical systems. \\\hline
	\caption{Resources for Numerical and Quantum Simulations.}
\end{longtable}
\end{center}
\begin{multicols}{2}
\section{Concluding Remarks}
In this paper, we have covered a handful of areas out of the vast versatile potential domains which can show quantum advantage towards quantum simulation in near future. Today, the real hardware implementation is limited to elementary quantum systems and processes which is due to limitation in realising decoherence free long circuit run time and inevitable gate errors. But with every new day, we have seen new algorithms and techniques coming up which have been enlightening the scientific community with optimised methods to realise quantum simulation and this only narrates that, realising the advantage of quantum simulation on quantum computers is a reality not very far now. 
The realisation of quantum treatment with Hamiltonian simulation has also expanded its branches to fundamental physics like- simulating gauge theories\cite{latticegauge}\cite{latticegauge2}, problems in high energy physics\cite{hep1}\cite{hep2}\cite{quantumscattering} and quantum sensing. Its only a matter of time when quantum computers will be solving real world problems addressing quantum simulations.
\section{Acknowledgement}
We are thankful to our colleagues and collaborators who contributed their expertise, feedback, and valuable insights during the preparation of this paper.

This work was made possible by the collective effort and unwavering commitment of our team at Qulabs, and we are immensely grateful for their collaboration and support.
\end{multicols}
\begin{multicols}{2}
\newpage
\bibliography{All_ref}
\end{multicols}
\end{document}